\documentclass[aps, reprint, onecolumn, superscriptaddress, nofootinbib]{revtex4-1}

\usepackage{amsmath,amssymb,amsfonts,physics,float}
\usepackage{booktabs} 
\usepackage[setpagesize=false]{hyperref}
\usepackage{slashed}
\allowdisplaybreaks[4]
\usepackage{graphicx}
\usepackage[labelformat=simple]{subcaption} 

\newcommand{\lag}{\mathcal{L}}

\newcommand{\rl}{\mathrm{L}}
\newcommand{\rr}{\mathrm{R}}

\usepackage{ulem}
\usepackage{color}
\definecolor{ar}{rgb}{1.0, 0.01, 0.24}
\definecolor{al}{rgb}{0.82, 0.1, 0.26}
\definecolor{ev}{rgb}{0.56, 0.0, 1.0}

\def\be{\begin{eqnarray}}
	\def\ee{\end{eqnarray}}

\begin{document}

\title{
Chiral symmetry restoration and hyperon suppression in neutron stars
}

\author{Bikai Gao}
\email{bikai@rcnp.osaka-u.ac.jp}
\affiliation{Research Center for Nuclear Physics (RCNP), Osaka University, Osaka 567-0047, Japan}

\date{\today}

\begin{abstract}
The ``hyperon puzzle'' remains a fundamental challenge in nuclear astrophysics. We investigate hyperon emergence in neutron star matter using the $SU(3)$ parity doublet model with chiral representation $(3,\bar{3}) + (\bar{3},3)$. This framework naturally incorporates chiral symmetry restoration and provides a systematic description of baryon masses in dense matter through the interplay between the chiral condensate and the chiral invariant mass $m_0$. We find that the hyperon onset density exhibits strong sensitivity to $m_0$: for $m_0 = 500$ MeV, hyperons first appear at $1.9n_0$ while for $m_0 \gtrsim 750$ MeV, hyperons emerge only above $5n_0$. This delayed onset arises from the weakened density dependence of baryon masses at larger $m_0$ values. When the hyperon onset density exceeds the expected quark-hadron transition range ($2$--$5n_0$), matter undergoes deconfinement before hyperons populate, avoiding the EoS softening while maintaining consistency with massive neutron star observations. Our results demonstrate that chiral dynamics provides a natural resolution to the hyperon puzzle without requiring ad hoc repulsive hyperon interactions.
\end{abstract}

\maketitle

\section{INTRODUCTION}

Neutron star (NS) represents one of the densest objects in the universe, with central densities reaching several times nuclear saturation density $n_0 \approx 0.16$ fm$^{-3}$. Under such extreme conditions, the composition of matter becomes fundamentally different and various exotic degrees of freedom may emerge, including hyperons~\cite{Glendenning:1991es, Vidana:2000ew,Schaffner-Bielich:2000nft,Weissenborn:2011ut,Oertel:2016bki} (baryons containing strange quarks), deconfined quark matter\cite{Weber:2004kj,Kurkela:2009gj,Holdom:2017gdc,Baym:2017whm}, kaon or pion condensates~\cite{Dautry:1979bk, Ellis:1995kz, Muto:2021jms,Muto:2025jaq}, and other exotic phases~\cite{McLerran:2007qj,Kojo:2009ha,Fujimoto:2023mzy,Gao:2024jlp,Gao:2025rbq}. Understanding which of these degrees of freedom appear and at what densities they emerge is essential for determining the equation of state (EoS) of dense matter and corresponding neutron star properties such as mass, radius, and cooling behavior.

Recent astrophysical observations have provided valuable insights into the properties of NSs~\cite{PhysRevLett.119.161101,LIGOScientific:2017ync,LIGOScientific:2018cki,Riley:2021pdl,Miller:2021qha,Raaijmakers:2021uju,Dittmann:2024mbo,Salmi:2024aum,Riley:2019yda, Vinciguerra:2023qxq}. These observations impose stringent constraints on theoretical models of dense matter~\cite{Celi:2025zmn,Imam:2025lut,Folias:2025zmu,Nanopoulos:2025elj, Zhang:2025rnf,Yang:2025iyv}, especially for those incorporate hyperons. When hyperons emerge in NS matter at baryon densities of roughly $2$--$3n_0$, they introduce additional degrees of freedom that tend to reduce the pressure, thereby softening the EoS. Such softening generally reduce the maximum mass of NSs to values below the observed $\sim2\,M_{\odot}$ threshold, giving rise to the so-called hyperon puzzle~\cite{Lonardoni:2014bwa,Gerstung:2020ktv,Li:2020dst}.

Various theoretical approaches have been developed to address the hyperon puzzle. These include introducing strong repulsive hyperon-hyperon interactions through vector meson exchange~\cite{Schaffner:1995th,Weissenborn:2011ut,Schulze:2011zza}, considering three-body forces involving hyperons~\cite{Vidana:2010ip,Yamamoto:2014jga,Chatterjee:2015pua,Togashi:2016fky, Gerstung:2020ktv}, or invoking early phase transitions to quark matter before hyperons can appear~\cite{Baldo:1999rq, Klahn:2013kga,Kojo:2014rca}. However, these conventional approaches face several limitations. First, most of these models do not explicitly incorporate the chiral symmetry of QCD, which is a fundamental symmetry expected to play an important role. Second, the coupling constants between hyperons and mesons are often treated as free parameters that must be adjusted to fit observational constraints, leading to much uncertainties in predictions. Third, these models often struggle to simultaneously describe both vacuum baryon properties and high-density NS matter within a single consistent framework.

The parity doublet model~\cite{Detar:1988kn,Jido:2001nt} offers an alternative framework that addresses many of the limitations of conventional approaches by  incorporating chiral symmetry. In this model, baryons are organized into chiral multiplets in which positive- and negative-parity states form chiral partners. As the baryon density increases and chiral symmetry is restored, the masses of these parity partners become degenerate~\cite{Zschiesche:2006zj, Dexheimer:2007tn, Sasaki:2010bp, Sasaki:2011ff,Steinheimer:2011ea, Motohiro:2015taa, Mukherjee:2016nhb,  Gallas:2011qp, Suenaga:2017wbb,Marczenko:2017huu,  Marczenko:2018jui, Yamazaki:2019tuo, Minamikawa:2021fln,Yuan:2025dft,Kong:2025dwl,Gao:2025nkg,Gao:2025vdc}. This framework provides several key advantages: it preserves the chiral structure of QCD, naturally connects vacuum and high-density regimes through the density dependence of the chiral condensate, and reduces the number of free parameters by constraining coupling constants via the underlying symmetry relations. Moreover, the chiral structure restricts certain interactions while determining the allowed forms of others. The parity doublet model can be implemented through two distinct realizations of chiral symmetry. In the nonlinear realization, chiral symmetry acts nonlinearly on the baryon fields, following the traditional framework of chiral perturbation theory~\cite{Papazoglou:1998vr,Mishra:2003tr,Steinheimer:2011ea,Dexheimer:2012eu,Fraga:2023wtd}. This approach works well at low densities where chiral symmetry is strongly broken. In contrast, the linear realization treats chiral partners as explicit dynamical degrees of freedom with chiral symmetry acting linearly on the fields~\cite{Nishihara:2015fka,Minamikawa:2023ypn,Gao:2024mew}. This linear approach becomes particularly advantageous near and beyond the chiral restoration scale, as it provides a more natural and simpler framework for incorporating the full baryon octet along with their parity partners.

In this paper, we construct an $SU(3)$ parity doublet model using the linear realization to study hyperon appearance in NS matter. We focus on the chiral representation $(3,\bar{3}) + (\bar{3},3)$, which naturally emerges from the quark-diquark picture where baryons contain ``good'' (flavor-antisymmetric) diquarks. We systematically build the chiral effective Lagrangian including nucleons and hyperons ($\Lambda$, $\Sigma$, $\Xi$) along with their negative parity partners. Scalar and pseudoscalar mesons are introduced through a $3 \times 3$ matrix field, while vector mesons are incorporated using the hidden local symmetry~\cite{Bando:1987br, Harada:2000kb, Harada:2003jx} . We determine model parameters by fitting to vacuum baryon masses and nuclear matter saturation properties. Our main finding is that the chiral invariant mass $m_0$ plays a crucial role in controlling the onset density of hyperons. For sufficiently large values of $m_0$, hyperons only appear at densities where quark-hadron phase transitions are expected to occur ($\sim 5n_0$), potentially resolving the hyperon puzzle by preventing the EoS softening associated with hyperons.

This paper is organized as follows. In Section~\ref{sec_SU(2)}, we briefly review the $SU(2)$ parity doublet model to establish notation and basic concepts. Section~\ref{sec_SU(3)} presents the construction of the $SU(3)$ parity doublet model, including the choice of chiral representations, the effective Lagrangian, and the treatment of explicit symmetry breaking to reproduce the observed baryon mass spectrum. Section~\ref{sec_Numerical} describes the numerical analysis, including parameter determination and the calculation of hyperon onset densities for different values of the chiral invariant mass. Finally, Section~\ref{summary} provides a summary and discusses future directions.

\section{Brief review of SU(2) parity doublet} \label{sec_SU(2)}
In the simple $SU(2)$ parity doublet model~\cite{Zschiesche:2006zj,Dexheimer:2007tn,Motohiro:2015taa,Gao:2025nkg,Gao:2025vdc,Recchi:2025pyy}, the ordinary nucleon field $N_1$ with positive parity is regarded as the chiral partner to the one with negative parity $N_2$ with their chiral representations for left- handed and right- handed are 
\begin{equation}
\begin{aligned}
N_{1L} \sim (2, 1)&, \quad N_{1R} \sim (1, 2) \\
N_{2L} \sim (1, 2)&, \quad N_{2R} \sim (2, 1).
\end{aligned}
\end{equation}
Under chiral transformation, they transform as
\begin{equation}
\begin{aligned}
N_{1L} \rightarrow g_L N_{1L},& \quad N_{1R} \rightarrow g_R N_{1R} \\
N_{2L} \rightarrow g_R N_{2L},& \quad N_{2R} \rightarrow g_L N_{2R}.
\end{aligned}
\end{equation}
We  then introduce the meson field with linear representation $M = \sigma + i \pi^a \tau_a$ and the corresponding chiral transformation property as
\begin{align}
M \rightarrow  g_L M g_R^\dagger. 
\end{align}
Here the $\sigma$ denotes for the iso-singlet scalar field, $\pi$ is the pion field and $\tau$ the Pauli matrices. Using these fields, the chiral invariant Lagrangian can be constructed 
\begin{equation}
\begin{aligned}
\mathcal{L} =& \bar{N}_1 i \slashed{\partial} N_1 - g_1 \bar{N}_1 M N_1   \\
 &+ \bar{N}_2 i \slashed{\partial} N_2  - g_2\bar{N}_2 M^{\dagger} N_2  \\
 & - m_0 (\bar{N}_1\gamma_5 N_2 - \bar{N}_2 \gamma_5 N_1) + \mathcal{L}_{{\rm meson}}
\end{aligned}
\end{equation}
with the mass matrix calculated as
\begin{align}
\hat{M}  = 
\left[
\begin{matrix}
g_1\sigma & m_0 \gamma_5 \\
- m_0 \gamma_5 & g_2 \sigma \\
\end{matrix}\right]
\end{align}
This mass matrix can be diagonalized  with the mixing angle
\begin{align}
{\rm tan} 2\theta = \frac{2 m_0}{ (g_1 + g_2) \sigma}.
\end{align}
and the corresponding masses of two states are given as
\begin{align}
m_\pm = \frac{1}{2} \left( \sqrt{ (g_1 + g_2)^2\sigma^2 + 4m_0^2 }  \pm (g_1 - g_2)\sigma \right).
\end{align}
After chiral symmetry is restored $\sigma = 0$, we can easily see the mass of two nucleon states become degenerate to $m_0$. We can rewrite the mass formula to obtain $g_1, g_2$ values from the vacuum inputs with $m_+ = 939$ MeV and $m_-= 1535$ MeV
\begin{equation}
g_{1,2}=\frac{1}{2 f_\pi}\left(\sqrt{\left(m_{-}+m_{+}\right)^2-4 m_0^2} \pm\left(m_{-}-m_{+}\right)\right).
\end{equation}
At vacuum,  we take the expectation value of $\sigma$ to be the pion decay constant $f_\pi$ and then corresponding determined $g_1, g_2$ are listed in Table.~\ref{table_SU2}.
\begin{table} [tb]
\centering
\caption{Values of coupling constants $g_1, g_2$ for different choices of chiral invariant mass $m_0$.}
\begin{tabular}{c|ccccc}
\hline \hline$m_0[\mathrm{MeV}]$ ~&~ 500 ~&~ 600 ~&~ 700 ~&~ 800 ~&~ 900~\\
\hline
$g_1$ & 15.47 & 14.93 & 14.26 & 13.44 & 12.41  \\
$g_2$ & 9.02 & 8.48 & 7.81 & 6.99  & 5.96 \\
\hline \hline
\end{tabular}\label{table_SU2}
\end{table}

\section{SU(3) parity doublet model}\label{sec_SU(3)}
In this section, we will construct the $SU(3)$ parity doublet model begin from the chiral representation of quarks and discuss about the possible chiral representation for baryons. Using the proper chiral representation, we construct the Lagrangian based on $SU(3)_L \times SU(3)_R \times U(1)_A$ symmetry~\cite{Nishihara:2015fka,Chen:2009sf,Chen:2010ba,Gao:2022klm} with additional explicit symmetry breaking term~\cite{Schechter:1969kn,Papazoglou:1997uw,Fraga:2023wtd} to reproduce the hyperon mass spectrum. Finally, we construct the thermodynamic potential and discuss about the onset density of hyperon inside neutron star.

\subsection{Chiral representations for baryons}\label{sec-representation}
Quarks transform under $SU(3)_L \times SU(3)_R \times U(1)_A$ symmetry as
\begin{equation}
\begin{aligned}
&q_{L} \rightarrow  e^{-i \theta_{A}} g_{L}q_{L}, \\
&q_{R} \rightarrow  e^{+i \theta_{A}} g_{R}q_{R},
\end{aligned}
\end{equation}
with $g_{L,R} \in$ $SU(3)_{L,R}$ and $\theta_{A}$ being the $U(1)_A$ transformation parameters. 
Accordingly, we assign the U(1)$_{A}$ charge of the left and right handed quarks as $-1$ and $+1$, respectively. The chiral representation of the left and right handed quark is then given by
\begin{equation}
q_{L} :( 3, 1 )_{-1}, \quad q_R :(1 , 3 )_{+1}
\end{equation}
where these $3$ and $1$ in the bracket express the triplet and singlet for $SU(3)_{L}$ symmetry and $SU(3)_{R}$ symmetry, respectively. 
The index indicates the axial charge of the fields. 

Since baryons consist of three valence quarks, 
the baryon fields are related with the tensor products of three quark fields. 
We define the left-handed baryon field 
as a product of a spectator left-handed quark 
and left- or right-handed diquark, 
while the right-handed baryon has a right-handed spectator quark. 
Taking irreducible decomposition, the left-handed baryon 
can be expressed as the following representations 
\begin{align}
q_\rl&\otimes(q_\rl\otimes q_\rl+q_\rr\otimes q_\rr)\notag\\
\sim&(1,1)_{-3}+(8,1)_{-3}+(8,1)_{-3}\notag\\
&+(10,1)_{-3} +(3,\bar3)_{+1}+(3,6)_{+1}\,. 
\end{align}
After the chiral symmetry is spontaneously broken down to flavor $SU(3)_F$ symmetry, octet baryons appear from the representations of $(3,\bar3)$, $(8,1)$, and $(3,6)$. The representations $(3,\bar3)$  contain flavor-antisymmetric diquarks $\sim\bar3$ which is called ``good'' diquark,  
while $(3,6)$ contains flavor-symmetric diquark $\sim6$ called ``bad'' diquark. Also, for simplicity, we only focus on baryons having spin 1/2. 

The diquark picture provides essential insights into baryon structure and dynamics. While baryons fundamentally consist of three valence quarks, the complexity of the three-body problem in QCD makes direct calculations extremely challenging. The emergence of diquark correlations offers a crucial simplification: the intricate three-quark system reduces to a more tractable quark-diquark configuration, where two quarks form a correlated pair that interacts with the third spectator quark.
This reduction reflects important physical dynamics rather than mere mathematical convenience. The one-gluon exchange interaction creates an attractive force in the color-antitriplet channel, naturally favoring diquark formation. Among these correlations, ``good'' diquarks in the flavor-antisymmetric $\bar{3}$ representation are particularly favored---they benefit from both attractive color forces and spin-flavor correlations, resulting in compact, tightly bound configurations. In contrast, ``bad'' diquarks with symmetric flavor structure in the $6$ representation experience weaker binding despite the same color attraction.

These different binding strengths directly manifest in the baryon mass spectrum. Baryons dominated by good diquark correlations systematically exhibit lower masses than those containing bad diquarks, explaining key features of the mass hierarchy within baryon multiplets. This observation has profound implications for the choice of chiral representations in effective theories.
Given that ground state baryons are energetically favored to contain good diquarks in the $\bar{3}$ representation, the chiral representation $(3, \bar{3}) + (\bar{3}, 3)$ emerges as the natural and optimal framework for describing ground state hyperons and their chiral partners. This representation directly incorporates the dominant diquark correlations, ensuring that the effective theory captures the essential physics. Therefore, in the following sections, we construct the chiral effective Lagrangian based on the $(3, \bar{3}) + (\bar{3}, 3)$ representations.

\subsection{Parity doublet structure of baryons}\label{sec_3_1}
Let us  consider the  model including the (3, $\bar{3}$) + ($\bar{3}$, 3) representations for octet baryons at the leading order of $M$. We introduce the following fields:
\begin{equation}
\begin{aligned}
\psi_{1L} \sim (3, \bar{3})_{+1}, \quad &\psi_{1R} \sim (\bar{3}, 3)_{-1}, \\
\psi_{2L} \sim (\bar{3}, 3)_{-1}, \quad &\psi_{2R} \sim (3, \bar{3})_{+1}, \\
\end{aligned}
\end{equation}
The fields $\psi_1$ and $\psi_2$ form chiral partners with opposite assignments of the chiral representations---a ``mirror'' assignment similar to $SU(2)$ case. Under chiral transformation, these fields transform as
\begin{equation}
\begin{aligned}
\psi_{1L} \rightarrow g_L \psi_{1L} g_{R}^{\dagger}, \quad \psi_{1R} \rightarrow g_{R}\psi_{1R} g_{L}^{\dagger},\\
\psi_{2L} \rightarrow g_{R}\psi_{2L} g_{L}^{\dagger}, \quad \psi_{2R} \rightarrow g_L \psi_{2R} g_{R}^{\dagger},
\end{aligned}
\end{equation}

The explicit component of the $\psi$ field is 
\begin{align}
(\psi)^i_j&\sim
\frac1{\sqrt3}\Lambda_0 + (B)^i_j \\
(B)^i_j &=\begin{bmatrix}
\frac1{\sqrt2}\Sigma^0+\frac1{\sqrt6}\Lambda & \Sigma^+ & p \\
\Sigma^- & -\frac1{\sqrt2}\Sigma^0+\frac1{\sqrt6}\Lambda & n \\
\Xi^- & \Xi^0 & -\frac2{\sqrt6}\Lambda \\
\end{bmatrix}\,, \label{eq_octet}
\end{align}
for left-handed and right-handed respectively.
We also introduce a $3 \times 3$ matrix field $M$ for scalar and pseudo-scalar mesons as
\begin{align}
M \sim (3 , \bar{3})_{-2}.
\end{align}
This field is made from $(q_L\bar{q}_R)$ and have the chiral transformation property as
\begin{align}
M \rightarrow g_L M g_{R}^{\dagger}.
\end{align}
By taking the mean-field approximation, the meson field is expressed as
\begin{align}\label{eq_mean_field}
\langle M \rangle = {\rm diag}(\sigma_0, \sigma_0, \sigma_{s0}).
\end{align}
At vacuum $\sigma_0= f_\pi$ and to incorporate explicit flavor symmetry breaking effects, we choose $\sigma_{s0} = 2f_K - f_\pi$, where  $f_K$ is the kaon decay constants with the explicit values shown in Table.~\ref{tab-condensate-input}. These parameter choices naturally encode the mass difference between strange and non-strange quarks through the meson field expectation values.
\begin{table}
\caption{
Physical inputs of the decay constants for pion and kaon,
and the VEV of the meson field  $\langle M \rangle={\rm diag}\{{\sigma_0,\sigma_0,\sigma_{s0}}\}$. 
}
\label{tab-condensate-input}
\centering
\begin{tabular}{c|c}
\hline\hline
$f_\pi$ & 92.1 MeV \\
$f_K$ & 110 MeV \\
$\sigma_0$ & $f_\pi(=92.1\,\mathrm{MeV})$ \\
$\sigma_{s0}$ & $2f_K-f_\pi(=128\,\mathrm{MeV})$ \\
\hline\hline
\end{tabular}
\end{table}

With these configurations, the leading-order chiral invariant Yukawa interactions can be constructed based on $SU(3)_L \times SU(3)_R \times U(1)_A$ symmetry as

\begin{align}
\mathcal{L}_{B} = \mathcal{L}_{{\rm kin}} + \mathcal{L}_{{\rm CIM}} + \mathcal{L}_{{\rm diag}} ,
\end{align}
with the kinetic term 
\begin{align}
\mathcal{L}_{{\rm kin}} =& {\rm tr}(\bar{\psi}_{1L}i \slashed{D} \psi_{1L}) + {\rm tr}(\bar{\psi}_{1R}i \slashed{D} \psi_{1R}) \nonumber\\
&+{\rm tr}(\bar{\psi}_{2L}i \slashed{D} \psi_{2L}) + {\rm tr}(\bar{\psi}_{2R}i \slashed{D} \psi_{2R}).
\end{align}
Here the covariant derivatives for each field are 
 \begin{equation}\label{eq_kine}
 \begin{aligned}
 D_{\mu}\psi_{1L, 2R} =& \partial_\mu \psi_{1L, 2R} - i \mathcal{L}_\mu \psi_{1L, 2R} + i \psi_{1L, 2R}\mathcal{R}_\mu, \\
 D_{\mu}\psi_{1R, 2L} =& \partial_\mu \psi_{1R, 2L} - i \mathcal{R}_\mu \psi_{1R, 2L} + i \psi_{1R, 2L}\mathcal{L}_\mu, 
 \end{aligned}
 \end{equation}
 where the $\mathcal{L}_\mu$ and $\mathcal{R}_\mu$ represent the external gauge fields arising from the gauging of the chiral $SU(3)_L \times SU(3)_R$ symmetry.

The chiral invariant mass term is expressed as
\begin{align}\label{eq_cim}
\mathcal{L}_{{\rm CIM}} =&  m_0 \left[ {\rm tr}(\bar{\psi}_{1R}  \psi_{2L}) + {\rm tr}(\bar{\psi}_{1L} \psi_{2R})\right] .
\end{align}
The diagonal Yukawa interactions are given by
\begin{align}
\lag_\mathrm{diag}=&  g_1\left[
\varepsilon_{r_1r_2r_3}\varepsilon^{l_1l_2l_3}
(\bar\psi_{1L})^{r_1}_{l_1}(M^\dagger)^{r_2}_{l_2}(\psi_{1R})^{r_3}_{l_3}
+\mathrm{h.c.}\right] \nonumber \\
&+g_2\left[
\varepsilon_{l_1l_2l_3}\varepsilon^{r_1r_2r_3}
(\bar\psi_{2L})^{l_1}_{r_1}(M)^{l_2}_{r_2}(\psi_{2R})^{l_3}_{r_3}
+\mathrm{h.c.}\right] \label{eq_g1g2}
\end{align}

For simplicity, we also define the isospin vectors as 
\begin{equation}
\begin{aligned}
\psi_N&\equiv(\psi_p,\psi_n)\\
\psi_\Sigma&\equiv(\psi_{\Sigma^-},\psi_{\Sigma^0},\psi_{\Sigma^+})\\
\psi_\Xi&\equiv(\psi_{\Xi^-},\psi_{\Xi^0})\,. 
\end{aligned}
\end{equation}
Then for each baryon field 
\begin{align}
\Psi_H = (\psi_{1H}, \gamma_5 \psi_{2H})^{T}, \quad (H = N,\Lambda, \Sigma, \Xi),
\end{align}
we can write the mass term as
\begin{align}
\mathcal{L}_{{\rm mass}} = - \sum_{H = N, \Lambda, \Sigma, \Xi} \bar{\Psi}_H \hat{M}_H \Psi_H,
\label{eq_mass_L}
\end{align} 
the corresponding mass matrix for each baryon is calculated as
\begin{align}
&\hat{M}_N  (g_1, g_2) = \left(\begin{matrix}
g_1 \sigma_0  & m_0  \\
m_0 & -g_2 \sigma_0  \\
\end{matrix} \right) \label{eq_mass_N}\\
&\hat{M}_\Sigma  (g_1, g_2) =\left(\begin{matrix}
g_1 \sigma_{s0}  & m_0 \\
 m_0  & -g_2 \sigma_{s0} \\
\end{matrix} \right)\\
&\hat{M}_\Xi (g_1, g_2)=  \left(\begin{matrix}
g_1 \sigma_0  & m_0\\
m_0   &-g_2 \sigma_0  \\
\end{matrix} \right) \\
&\hat{M}_\Lambda (g_1, g_2) = \left(\begin{matrix}
\frac{1}{3}g_1 (4 \sigma_0 - \sigma_{s0})  & m_0 \\
m_0  & -\frac{1}{3}g_2( 4 \sigma_0 - \sigma_{s0}) \\
\end{matrix} \right) \label{eq_mass_Lam}, 
\end{align}
where $\sigma_0$ and $\sigma_{s0}$ denote the expectation value of $\sigma$ and $\sigma_s$ in vacuum respectively. After diagonal the mass matrix, we obtain the physical masses for each baryon species
\begin{equation}\label{eq_mass_pm}
\begin{aligned}
    \hat{m}_{N, \pm} &= \sqrt{m_0^2 + (\frac{g_1 + g_2}{2})^2\sigma_0^2} \mp \frac{g_1 - g_2}{2}\sigma_0, \\
    \hat{m}_{\Sigma, \pm} &= \sqrt{m_0^2 + (\frac{g_1 + g_2}{2})^2\sigma_{s0}^2} \mp \frac{g_1 - g_2}{2}\sigma_{s0}, \\
     \hat{m}_{\Xi, \pm} &= \sqrt{m_0^2 + (\frac{g_1 + g_2}{2})^2\sigma_0^2} \mp \frac{g_1 - g_2}{2}\sigma_0, \\
    \hat{m}_{\Lambda, \pm} &= \sqrt{m_0^2 + (\frac{g_1 + g_2}{2})^2(\frac{4 \sigma_0 - \sigma_{s0}}{3})^2} \mp \frac{g_1 - g_2}{2}(\frac{4 \sigma_0 - \sigma_{s0}}{3}),
\end{aligned}
\end{equation}
In this form, the nucleon have the same mass with the $\Xi$ baryon. To fix the vacuum mass, we incorporate the explicit symmetry breaking terms~\cite{Schechter:1969kn,Papazoglou:1997uw,Fraga:2023wtd} in the hypercharge direction with the Lagrangian written as
\begin{align}
\mathcal{L}_{ {\rm explicit}} = - m_1 {\rm tr}\left( \bar{\Psi}\Psi - \bar{\Psi}\Psi S \right) - m_2 {\rm tr}(\bar{\Psi}S \Psi),
\end{align}
where $S = {\rm diag}(0, 0, 1)$. The first term, ${\rm tr}(\bar{\Psi}\Psi)$, preserves both chiral and flavor symmetries, similar to the chiral-invariant mass term in Eq.~(\ref{eq_cim}). In contrast, the second and third terms, ${\rm tr}(\bar{\Psi}\Psi S)$ and ${\rm tr}(\bar{\Psi}S\Psi)$, break both chiral and flavor symmetries. Compare with previous work~\cite{Gao:2025eax} where all the terms are chiral invariant, these chiral breaking terms introduce new degrees of freedom and different approach to fix the hyperon masses.
Then the baryon mass for each species become
\begin{equation}
    \begin{aligned}
        m_{N, \pm} &= \hat{m}_{N, \pm}, \\
        m_{\Sigma, \pm} &= \hat{m}_{\Sigma, \pm} + m_1,\\
        m_{\Xi, \pm} & = \hat{m}_{\Xi, \pm} +m_1 + m_2,\\
        m_{\Lambda, \pm}&=\hat{m}_{\Lambda, \pm}+ \frac{m_1 + 2m_2}{3}.
    \end{aligned}
\end{equation}
The baryonic part of the model contains four free parameters: $g_1$, $g_2$, $m_1$, and $m_2$. To determine these parameters, we use the following experimentally measured baryon masses from the PDG ~\cite{ParticleDataGroup:2024cfk} as input:
\begin{equation}
\begin{aligned}
m_{N,+} = 939\,{\rm MeV},& \quad m_{N, -}= 1535\,{\rm MeV}, \\
m_{\Lambda, +} = 1116\,{\rm MeV},& \quad m_{\Sigma, +}=1193\,{\rm MeV}.
\end{aligned}
\end{equation}
The ground state nucleon and its excited state $N^*(1535)$ form a chiral partner pair. Since the chiral partners of the ground state $\Lambda$, $\Sigma$, and $\Xi$ baryons are not yet well-established, we treat their masses as predictions of our model.
\begin{figure}[htbp]
\centering
\includegraphics[width=0.5\hsize]{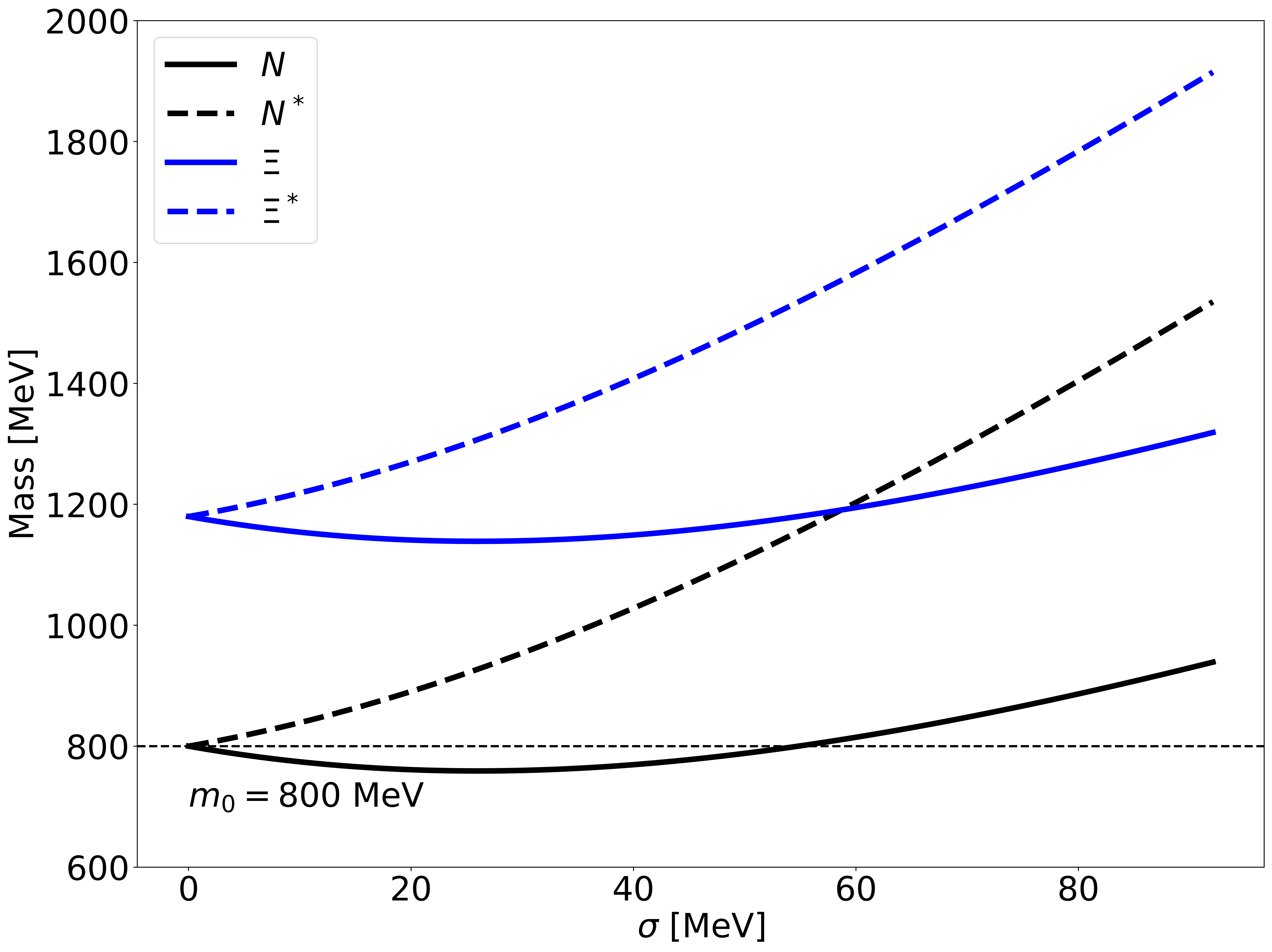}
\caption{Masses of $N$ and $\Xi$ as functions of the chiral condensate $\sigma$, with $m_0$ fixed to be 800~MeV.  }
\label{fig_mass_sigma}
\end{figure}
We note that alternative choices of input masses are possible. For instance, one could use different combinations such as $\Lambda$ and $\Xi$, or $\Sigma$ and $\Xi$ ground states along with the nucleon states. We have verified that such alternative parametrizations yield consistent mass spectra with only minor variations in the predicted values. 

For a range of chiral invariant masses $m_0 = 500$--900~MeV, the fitted parameters $g_1$, $g_2$, $m_1$, and $m_2$ are listed in Table~\ref{tab:coupling_constant}. The corresponding baryon mass spectrum in vacuum is presented in Table~\ref{tab:baryon_masses}, where input values are shown in regular font and predicted values in bold. Our results successfully reproduce the input masses and provide predictions for the excited states. Notably, the predicted $\Xi$ ground state mass is in good agreement with the experimental value of $\Xi_{\rm G.S.} = 1318$~MeV from the PDG~\cite{ParticleDataGroup:2024cfk}. For the chiral partners, we predict: (i) the $\Lambda$ chiral partner at 1636.72~MeV, which can be associated with the observed $\Lambda^*(1670)$ resonance; (ii) the $\Xi$ chiral partner at approximately 1918~MeV, consistent with $\Xi^*(1950)$; and (iii) the $\Sigma$ chiral partner at 2014.85~MeV. While this last prediction appears somewhat higher than typical excited state masses, it remains possible given that multiple experimental resonances in this region could potentially correspond to chiral partners of the baryon octet.

In Fig.~\ref{fig_mass_sigma}, we take $m_0 = 800$~MeV as an example and plot the masses of $N$ and $\Xi$ since they are only the functions of the chiral condensate $\sigma$ as in Eq.~(\ref{eq_mass_pm}). As $\sigma$ decreases toward zero, chiral symmetry is gradually restored, and the masses of the ground states $N~(\Xi)$ and their excited partners $N^*~(\Xi^*)$ become degenerate. For the nucleon sector, the degenerate mass approaches the chiral-invariant mass $m_0$. In contrast, for the $\Xi$ sector, due to the presence of additional explicit symmetry-breaking terms $m_1$ and $m_2$, the degenerate mass approaches $m_0 + m_1 + m_2$. We note that the $\Sigma$ baryon mass depends only on the strange condensate $\sigma_s$ because the present framework includes only the $(3,\bar{3}) + (\bar{3},3)$ chiral representation. In a more complete treatment that also incorporates higher representations such as $(3,6) + (6,3)$, the $\Sigma$ baryon would acquire a dependence on the non-strange condensate $\sigma$, while the $\Xi$ baryon would also become sensitive to $\sigma_s$. Consequently, the overall dependence of each baryon mass on $\sigma$ and $\sigma_s$ would be modified.

\begin{table}[h]
\caption{Values of parameters $g_1, g_2, m_1, m_2$ for $m_0=500$--$900$ MeV.}
\label{tab:coupling_constant}
\centering
\begin{tabular}{lccccc}
\hline\hline
~$m_0$ [MeV] ~&~~ 500 ~~&~~ 600 ~~&~~~ 700 ~~~& ~~~800 ~~~&~~~ 900 ~\\
\hline
$g_1$ & 15.52 & 14.98 & 14.31 & 13.48 & 12.45 \\
$g_2$ & 9.05 & 8.51 & 7.84 & 7.01 & 5.98 \\
$m_1$ [MeV] & -34.43 & -3.88 & 32.99 & 76.63 & 127.65 \\
$m_2$ [MeV] & 420.07 & 387.27 & 348.34 & 303.19 & 251.60 \\
\hline\hline
\end{tabular}
\end{table}

\begin{table*}[htbp]
\centering
\caption{Summary table of all baryon masses with all the unit in [MeV]. The masses of $N(939), N(1535),\Lambda(1116)$ and $\Sigma(1193)$ are the input values and $\Lambda^*, \Sigma^*, \Xi(1318),\Xi^*$ are the predicted values shown in bold characters.   }
\label{tab:baryon_masses}
\begin{tabular}{l|ccccccccc}
\hline
\hline
$m_0$ ~~&~  $N(939)$ ~&~ $N^*(1535)$ ~&~ $\Lambda$(1116) ~&~ ${\bf \Lambda^*}$ ~&~ $\Sigma$(1193) ~&~ ${\bf \Sigma^*}$ ~&~ ${\bf \Xi}$(1318) ~&~ ${\bf \Xi^*}$ \\
\hline
500        & 939.00             & 1535.00                 & 1116.00 & 1636.72  &  1193.00 & 2014.85 & 1324.64 & 1920.64 \\
600    & 939.00   & 1535.00    & 1116.00 & 1636.72 & 1193.00 & 2014.85 & 1322.38 & 1918.38 \\
700 & 939.00  & 1535.00  & 1116.00 & 1636.72 & 1193.00  & 2014.85  & 1320.33 & 1916.33 \\
800 & 939.00   & 1535.00  & 1116.00  & 1636.72 & 1193.00 & 2014.85 & 1318.82 & 1914.82 \\
900 & 939.00      & 1535.00      & 1116.00 & 1636.72 & 1193.00 & 2014.85 & 1318.25 & 1914.25 \\
\hline
\hline
\end{tabular}
\end{table*}

\subsection{Scalar and psedoscalar mesons}
After fixing the baryon part, we turn to the meson sector with the scalar part of the Lagrangian as
\begin{align}
\mathcal{L}_{M}^{s} = \mathcal{L}_M^{{\rm kin}} - V_M - V_{{\rm SB}},
\end{align}
where
\begin{equation}
\begin{aligned}
\mathcal{L}_M^{{\rm kin}} &= \frac{1}{4}{\rm tr}[ \partial_\mu M \partial^\mu M^{\dagger} ], \\
V_M & = -\frac{1}{4}\bar{\mu}^2 {\rm tr}[ M M^{\dagger}] + \frac{1}{8} \lambda_4 {\rm tr}[ (MM^{\dagger})^2] \\
&\quad - \frac{1}{12}\lambda_6 {\rm tr}[ (M M^{\dagger})^3 ] + \lambda_8 {\rm tr}[ (MM^\dagger)^4 ], \\
V_{{\rm SB}} & = - \frac{1}{2} c {\rm tr}[\mathcal{M}^\dagger M + \mathcal{M} M^{\dagger}].
\end{aligned}
\end{equation}
Here $c$ is the coefficient for the explicit chiral symmetry breaking term from the bare quark matrix $\mathcal{M} = {\rm diag}\{ m_u, m_d, m_s\}$. We include terms with only one trace in $V_M$, which are expected to be the leading order in the $1 / N_c$ expansion. After taking the mean-field approximation as in Eq.~(\ref{eq_mean_field}), the Lagrangian become
\begin{equation}
\label{eq_Meson_three}
\begin{aligned}
\mathcal{L}_M^{{\rm kin}} &= \frac{1}{4} (2 \partial_{\mu} \sigma \partial^\mu \sigma^\dagger  + \partial_\mu \sigma_s \partial^\mu \sigma_s^{\dagger}), \\
V_M & = -\frac{1}{4}\bar{\mu}^2 (2 \sigma^2 + \sigma_s^2)+ \frac{1}{8} \lambda_4 (2\sigma^4 + \sigma_s^4 ) \\
&\quad - \frac{1}{12}\lambda_6 (2\sigma^6 + \sigma_s^6) + \lambda_8 (2\sigma^8 + \sigma_s^8), \\
V_{{\rm SB}} & = - (2 c m_u \sigma + c m_s \sigma_s).
\end{aligned}
\end{equation}
with $c m_u$ and $c m_s$ fixed from the pion decay constant $f_\pi$ and kaon decay constant $f_K$
\begin{align}
2c m_u = m_\pi^2 f_\pi, \quad c (m_u + m_s) = m_K^2 f_K.
\end{align}
In this study, we take $\sigma_s$ to be constant. 
In general, both $\sigma$ and $\sigma_s$ should depend on the baryon density (chemical potential), 
and they may also mix through the $U(1)_A$ anomaly term arising from the Kobayashi--Maskawa--'t~Hooft interaction~\cite{Gao:2022klm}. 
However, in the low-density region where hyperons have not yet appeared or their number densities are much smaller than those of nucleons, 
treating $\sigma_s$ as a constant is a well-justified approximation. 
This is because $\sigma_s$ primarily couples to hyperons through terms proportional to their scalar densities, 
which are vanishingly small when the system is dominated by non-strange baryons. 
Consequently, the source term driving the in-medium modification of $\sigma_s$ is effectively absent, 
and the field remains close to its vacuum expectation value determined by the stationary condition of the potential $V(\sigma_s)$. 
Furthermore, linear density approximation  suggest that the strange quark condensate 
$\langle \bar{s}s \rangle$ exhibits only a weak density dependence due to the  small $\sigma_{sN}$ term from recent lattic QCD simulations~\cite{Yang:2015uis, Abdel-Rehim:2016won, Bali:2016lvx, Gubler:2018ctz,Yamanaka:2018uud}.
This supports the assumption that $\sigma_s$ stays close to its vacuum value in the density region where strange baryons are not abundant. 
In this regime, the dynamics of chiral symmetry restoration are therefore governed almost entirely by the non-strange condensate $\sigma$, 
while the strange condensate acts as a static background field. 
Fixing $\sigma_s$ at its vacuum value thus provides a reasonable approximation that simplifies the model 
 when the hyperon number density is relatively small. 
This treatment further allows us to explore the density region where hyperons begin to appear 
and to assess the impact of their gradual emergence on the in-medium properties of baryons.

\subsection{Vector meson from hidden local symmetry}

The Lagrangian for vector mesons is introduced based on the hidden local symmetry (HLS)~\cite{Bando:1987br, Harada:2000kb, Harada:2003jx} by decomposing the scalar and pseudoscalar mesons as
\begin{align}
M = \xi^\dagger_L \Phi \xi_R
\end{align}
with $\Phi$ is a hermitian field and $\xi_{L, R}$ are unitary matrix fields.  Under chiral symmetry and the HLS, these fields transform as
\begin{equation}
\begin{aligned}
\xi_L &\rightarrow h(x)\, \xi_L \,g_L^\dagger, \\
\xi_R &\rightarrow h(x) \,\xi_R\, g_R^{\dagger}, \\
\Phi &\rightarrow h(x)\,\Phi \,h(x)^\dagger.
\end{aligned}
\end{equation}
Using these $\xi$ fields, we can define the Maurer-Cartan 1-forms as
\begin{equation}
\begin{aligned}
\alpha_{\mu}^{\perp}&= \frac{1}{2 i } \left[ (D_\mu \xi_R )\xi^{\dagger}_R - (D_\mu \xi_L)\xi_L^\dagger     \right],  \\
\alpha_{\mu}^{\parallel} &= \frac{1}{2 i} \left[ (D_\mu \xi_R )\xi^{\dagger}_R + (D_\mu \xi_L)\xi_L^\dagger   \right].
\end{aligned}
\end{equation}
with the transformation properties as
\begin{equation}
\begin{aligned}
\alpha_\mu^{\perp}  &\rightarrow h(x) \alpha_\mu^{\perp} h(x)^\dagger, \\
\alpha_\mu^{\parallel}  &\rightarrow h(x) \alpha_\mu^{\parallel} h(x)^\dagger + \frac{1}{i} (\partial_\mu h(x)) h(x)^\dagger.
\end{aligned}
\end{equation}
Also the covariant derivatives for $\xi$ and $\Phi$ fields are 
\begin{equation}
\begin{aligned}
D_\mu \xi_L & =\partial_\mu \xi_L-i \hat{V}_\mu \xi_L+i \xi_L  \mathcal{L}_\mu \\
D_\mu \xi_R & =\partial_\mu \xi_R-i \hat{V}_\mu \xi_R+i \xi_R \mathcal{R}_\mu \\
D_\mu \Phi & =\partial_\mu \Phi -i \hat{V}_\mu \Phi +i \Phi \hat{V}_\mu
\end{aligned}
\end{equation}
where $\mathcal{L}_\mu$ and $\mathcal{R}_\mu$ are external gauge fields correspond to Eq.~(\ref{eq_kine}) to introduce the chemical potential and $\hat{V}_\mu$ is the gauge field for the HLS which transformed as
\begin{align}
\hat{V}_\mu \rightarrow h(x) \hat{V}_\mu h(x)^{\dagger}
\end{align}
Assuming $\xi_L = \xi_R =1$, we obtain the vector part of the Lagrangian for baryons
\begin{equation}\label{eq_vector_B_L}
\begin{aligned}
\mathcal{L}_B^{{\rm Vec}} &= a_{V} {\rm tr} \left[ \bar{\psi}_{1L} \gamma^\mu (\xi^\dagger_L \hat{\alpha}_\mu^{\parallel} \xi_L) \psi_{1L} + \bar{\psi}_{1R} \gamma^\mu (\xi^\dagger \hat{\alpha}_\mu^{\parallel}\xi_R )\psi_{1R}  \right]  \\
&\quad + a_{V}^\prime {\rm tr}\left[ \bar{\psi}_{1L} \gamma^\mu \psi_{1L} ({\rm tr}\,\xi^\dagger_R \hat{\alpha}_\mu^{\parallel} \xi_R   - \xi^\dagger_R \hat{\alpha}_\mu^{\parallel} \xi_R) \right.\\
&\quad  \quad + \left. \bar{\psi}_{1R} \gamma^\mu \psi_{1R}({\rm tr}\,\xi^\dagger_L\hat{\alpha}_\mu^{\parallel} \xi_L  - \xi^\dagger_L \hat{\alpha}_\mu^{\parallel} \xi_L)  \right] + {\rm Mirror} , \\
& = - g a_V {\rm tr}\left[ \bar{\psi}_{1L} \gamma_\mu V_L^\mu \psi_{1L} + \bar{\psi}_{1R} \gamma_\mu V_R^\mu \psi_{1R} \right] \\
& \quad  - g a_V^\prime {\rm tr} \left[ \bar{\psi}_{1L} \gamma_\mu \psi_{1L} ({\rm tr}\, V_R^\mu - V_R^\mu) \right.\\
&\quad \quad  + \left. \bar{\psi}_{1R} \gamma^\mu \psi_{1R} ({\rm tr}V_L^\mu - V_L^\mu) \right] + {\rm Mirror}.
\end{aligned}
\end{equation}
where $g$ is the coupling constant of the vector meson interactions and the flavor -nonet vector mesons $\hat{V}$ is assigned as
\begin{align}
V^{\mu} =\sum_{a=0}^8 \nu_a^\mu \frac{\lambda_a}{2} = \frac{1}{\sqrt{2}}\left[
\begin{matrix}
(\omega^\mu + \rho^\mu) / \sqrt{2} & & \\
 & (\omega^\mu - \rho^\mu) / \sqrt{2} &\\
 & & \phi^\mu
\end{matrix}
 \right]\nonumber
\end{align}
We take the (1, 1) component of $\hat{V}=\delta_\mu^0 {\rm diag}(1, 0, 0)V_{11}$  as an example (detailed calculations are shown in Appendix.~\ref{sec_ap1}), Eq.~(\ref{eq_vector_B_L}) become
\begin{equation}
\begin{aligned}
\mathcal{L}_B^{{\rm Vec}} &= -g a_{V} {\rm tr}\,\psi_1^\dagger {\rm diag}(1, 0, 0) \psi_1 \\
&\quad - g a^\prime_V {\rm tr} \, \psi_1^{\dagger} \psi_1 \left[ 1 - {\rm diag} (1, 0, 0)\right] \\
& =  -g a_{V}\left[p^{\dagger}_1 p_1+\left(\Sigma^+_1\right)^{\dagger} \Sigma^{+}_1+\frac{1}{2}\left(\Sigma^0\right)^{\dagger}_1 \Sigma^0_1 \right.\\
&\left.+\frac{1}{3}\left(\Lambda^0_1\right)^{\dagger} \Lambda^0_1+\frac{1}{6}\left(\Lambda_1\right)^{\dagger} \Lambda_1\right] V_{11} \\
& -g a_{V}^{\prime}\left[p^{\dagger}_1 p_1+\left(\Sigma^{+}_1\right)^{\dagger} \Sigma^{+}_1 +\frac{1}{2}\left(\Sigma^0_1\right)^{\dagger} \Sigma^0_1+\frac{2}{3}\left(\Lambda^0_1\right)^{\dagger} \Lambda^0_1 \right.\\
&+\left.\frac{5}{6}\left(\Lambda_1\right)^{\dagger} \Lambda_1 + n^{\dagger}_1 n_1+\left(\Xi^0_1\right)^{\dagger} \Xi^0_1\right] V_{11} \\
& -g\left(a_{V}-a_{V}^{\prime}\right)\left[\frac{1}{3 \sqrt{2}}\left(\Lambda^0_1\right)^{\dagger} \Lambda_1+\frac{1}{\sqrt{6}}\left(\Lambda^0_1\right)^{\dagger} \Sigma^0_1 \right.\\
&\left.+\frac{1}{2 \sqrt{3}}\left(\Lambda_1\right)^{\dagger} \Sigma^0_1+\text { h.c. }\right] V_{11} + {\rm Mirror} .
\end{aligned}
\end{equation}
In this study, we set $a_V = a_V^\prime$ and all the cross terms of $\Lambda$ and $\Sigma$ vanishes. This is natural since the number of each flavor is conserved, baryons of different flavor ($N, \Sigma, \Xi, \Lambda$) do not mix when we consider hyperon interactions. After doing the full calculation, we can then define the effective chemical potentials as
\begin{equation}\label{eq_chemical_potential_0}
\begin{aligned}
\left(
\begin{matrix}
\hat{\mu}_p \\
\hat{\mu}_n
\end{matrix}
 \right)
 &= \mu_B + \mu_Q \left(
\begin{matrix}
1\\
0
\end{matrix}
 \right) - \frac{3}{2}g_{\omega}a_V \omega - \frac{1}{2}g_{\rho} a_V \rho \left(
\begin{matrix}
+1\\
-1
\end{matrix}
 \right) , \\
 \left(
\begin{matrix}
\hat{\mu}_{\Sigma^+} \\
\hat{\mu}_{\Sigma^0}\\
\hat{\mu}_{\Sigma^-}
\end{matrix}
 \right)& = \mu_B + \mu_Q \left(
\begin{matrix}
+1 \\
0\\
-1
\end{matrix}
 \right) - \mu_S - g_{\omega} a_V \omega \\
 &\quad - g_{\rho} a_V \rho  \left(
\begin{matrix}
+1 \\
0\\
-1
\end{matrix}
 \right)  - \frac{1}{\sqrt{2}}g_{\phi} a_V \phi , \\
 \hat{\mu}_{\Lambda} &= \mu_B - \mu_S- g_{\omega} a_V \omega - \frac{1}{\sqrt{2}} g_{\phi} a_V \phi,\\
 \left(
\begin{matrix}
\hat{\mu}_{\Xi^0} \\
\hat{\mu}_{\Xi^-}
\end{matrix}
 \right) &= \mu_B +  \mu_Q \left(
\begin{matrix}
0 \\
-1
\end{matrix}
 \right) - 2 \mu_S - \frac{1}{2}g_{\omega} a_V \omega \\
 & \quad - \frac{1}{2}g_{\rho} a_V \rho\left(
\begin{matrix}
+1\\
-1
\end{matrix}
 \right) - \frac{2}{\sqrt{2}} g_{\phi} a_V \phi.
\end{aligned}
\end{equation}
where we use the baryon chemical potential $\mu_B$, the electromagnetic charge chemical potential $\mu_Q$, and the strangeness chemical potential $\mu_S$. If we use the isospin chemical potential $\mu_I$ instead of $\mu_Q$, the effective chemical potentials are 
\begin{equation}\label{eq_chemical_potential}
\begin{aligned}
\left(
\begin{matrix}
\hat{\mu}_p \\
\hat{\mu}_n
\end{matrix}
 \right)
 &= \hat{\mu}_B + \hat{\mu}_I \left(
\begin{matrix}
+1/2 \\
-1/2
\end{matrix}
 \right), \\
 \left(
\begin{matrix}
\hat{\mu}_{\Sigma^+} \\
\hat{\mu}_{\Sigma^0}\\
\hat{\mu}_{\Sigma^-}
\end{matrix}
 \right)& = \hat{\mu}_B + \hat{\mu}_I \left(
\begin{matrix}
+1 \\
0\\
-1
\end{matrix}
 \right) - \hat{\mu}_S, \\
 \hat{\mu}_{\Lambda} &= \hat{\mu}_B - \hat{\mu}_S,\\
 \left(
\begin{matrix}
\hat{\mu}_{\Xi^0} \\
\hat{\mu}_{\Xi^-}
\end{matrix}
 \right) &= \hat{\mu}_B + \hat{\mu}_I \left(
\begin{matrix}
+1/2 \\
-1/2
\end{matrix}
 \right) - 2 \hat{\mu}_S.
\end{aligned}
\end{equation}
with 
\begin{equation}\label{eq_chemical_2}
\begin{aligned}
\hat{\mu}_B &= \mu_B - \frac{3}{2}g_{\omega} a_V \omega\\
&= \mu_B - g_{\omega NN} \omega, \\
\hat{\mu}_I &= \mu_I - g_{\rho} a_V\rho \\
&= \mu_I - g_{\rho NN}\rho, \\
\hat{\mu}_S &= \mu_S + \frac{1}{\sqrt{2}} g_{\phi} a_V \phi  - \frac{1}{2}g_{\omega} a_V \omega \\
&= \mu_S + g_{\phi NN} \phi - \frac{1}{3} g_{\omega NN} \omega
\end{aligned}
 \end{equation}
with $\mu_B$, $\mu_I$, and $\mu_S$ representing the baryon, isospin, and strangeness chemical potentials, respectively. In NS matter, due to the $\beta$ equilibrium, we can simply take the $\mu_S$ to be zero and the couplings $g_{\omega NN}, g_{\rho NN}$ and $g_{\phi NN}$ are parameters remained to be determined. 


\subsection{Nuclear matter equation of state}

Following previous works in Ref.~\cite{Gao:2022klm}, we obtain the thermodynamic potential in our chiral effective hadronic model as
\begin{equation}
\begin{aligned}
\Omega_H =& V(\sigma, \sigma_s) - V(\sigma_0, \sigma_{s0})  - \frac{1}{2}m_\omega^2 \omega^2 - \frac{1}{2} m_\rho^2 \rho^2 - \frac{1}{2}m_\phi^2\phi^2  \\
&- \lambda_{\omega \rho} (g_{\omega NN} \omega)^2 (g_{\rho NN}\rho)^2 \\
&-2 \sum_H \sum_\alpha \sum_{i = \pm}  \int^{k^{i, \alpha }_H} \frac{d^3 p}{(2\pi)^3} (\hat{\mu}_{H}^{i, \alpha} - E_H^{i, \alpha}).
\end{aligned}
\end{equation}
Here the sum of $H$ run over each baryon species $N, \Sigma, \Xi, \Lambda$ and  $\alpha$ sum over the isospin. Also, $i = \pm$ denote for the spin and $k_H^{i, \alpha}$ is the Fermi momentum for each baryon species with the corresponding single-particle energy $E_{H}^{i, \alpha} = \sqrt{ (k_H^{i, \alpha})^2 + m^2_{H, i} }$.  $\hat{\mu}_H^{i, \alpha}$ is the effective chemical potential for each baryon species defined in Eq.~(\ref{eq_chemical_potential}) and Eq.~(\ref{eq_chemical_2}). We also note that the vector meson mixing $\omega^2\rho^2$ term is introduced to adjust the slope parameter with the coefficient $\lambda_{\omega \rho}$.
The potential $V(\sigma, \sigma_s)$ term is 
\begin{equation}
\begin{aligned}
V(\sigma, \sigma_s) &= V(\sigma) + V(\sigma_s), \\
V(\sigma) &= -\frac{1}{2}\bar{\mu}^2 \sigma^2   + \frac{1}{4}\lambda_4 \sigma^4  - \frac{1}{6}\lambda_6\sigma^6  + 2\lambda_8  \sigma^8   - 2 c m_u\sigma , \\
V(\sigma_s) &= -\frac{1}{4}\bar{\mu}^2 \sigma_s^2  + \frac{1}{8}\lambda_4 \sigma_s^4 - \frac{1}{12}\lambda_6 \sigma_s^6 + \lambda_8  \sigma_s^8   - cm_s \sigma_s .
\end{aligned}
\end{equation}
We note here that the higher order of $\lambda_8$ term is needed to fix the vacuum value  of $\sigma_s$ to be $2f_K - f_\pi$ through the stationary condition
\begin{align}\label{eq_sigmas}
\frac{\partial V(\sigma_s)}{\partial \sigma_s} = -\frac{1}{2}\bar{\mu}^2 \sigma_s + \frac{1}{2} \lambda_4 \sigma_s^3 - \frac{1}{2}\lambda_6 \sigma_s^5 +  8 \lambda_8 \sigma_s^7 - cm_s=0.
\end{align}
We subtracted the potential in vacuum $V(\sigma_0, \sigma_{s0})$ to ensure the total potential in vacuum is zero. Since we take $\sigma_s$ as constant so that $\sigma_s = \sigma_{s0}$,  after the subtraction, 
\begin{equation}
\hat{V} = V(\sigma, \sigma_s) - V(\sigma_0, \sigma_{s0}),
\end{equation}
 is then only the function of $\sigma$ with $\sigma_s$ part completely cancels out. The baryon number density, isospin number density and the strangeness number density is calculated as
\begin{equation}
\begin{aligned}
n_B &= -\frac{\partial \Omega}{\partial \mu_B} \\
&= \frac{1}{3 \pi^2} \sum_{i=\pm}   \left[  \sum_{\alpha= \pm}(k_N^{i, \alpha})^3  +\sum_{\alpha= 0, \pm}(k_\Sigma^{i, \alpha})^3\right. \\
&\quad \left.+ \sum_{\alpha=  \pm}(k_\Xi^{i, \alpha})^3 + (k_\Lambda^{i, \alpha})^3 \right] \\
& = \sum_{i=\pm} n_p^i + n_n^i + n_{\Sigma^+}^i+ n_{\Sigma^0}^i + n_{\Sigma^-}^i \\
&\quad + n_{\Xi^+}^i + n_{\Xi^-}^i  + n_{\Lambda}^i , \\
n_I & = -\frac{\partial \Omega }{\partial \mu_I} \\
& = \sum_{i = \pm} \frac{n_p^i - n_n^i}{2} + (n^i_{\Sigma^+} - n^i_{\Sigma^-}) \\
&\quad + \frac{n^i_{\Xi^0} - n^i_{\Xi^-}}{2}, \\
n_S & = - \sum_{i = \pm} \left[ n_{\Sigma^+}^i +n_{\Sigma^0}^i + n_{\Sigma^-}^i + n_{\Lambda}^i + 2 (n_{\Xi^+}^i + n_{\Xi^-}^i)   \right]
\end{aligned}
\end{equation}
Using the definition of the number density, we  obtain the gap equations for vector mesons as
\begin{equation}
\begin{aligned}
\frac{\partial \Omega_H}{ \partial \omega} &= - m_\omega^2 \omega + g_{\omega NN} n_B - 2\lambda_{\omega \rho}g_{\omega NN}g_{\rho NN} \omega \rho^2=0 \\
\frac{\partial \Omega_H}{ \partial \rho} &= - m_{\rho}^2 \rho + g_{\rho NN} n_I - 2\lambda_{\omega \rho}g_{\omega NN}g_{\rho NN} \omega^2 \rho=0, \\
\frac{\partial \Omega_H}{ \partial \phi} & = - m_{\phi}^2 \phi + g_{\phi NN} n_S=0.
\end{aligned}
\end{equation}
Then the mean field $\omega, \rho$ and $\phi$ can be obtained from 
\begin{equation}\label{eq_vector_gap}
\begin{aligned}
\omega & = \frac{g_{\omega NN} n_B}{ m_{\omega}^2 + 2\lambda_{\omega \rho} g_{\omega NN} g_{\rho NN}\rho^2 }, \\
\rho& = \frac{g_{\rho NN} n_I}{ m_{\rho}^2 + 2\lambda_{\omega \rho}g_{\omega NN} g_{\rho NN} \omega^2  }, \\
 \phi &= \frac{g_{\phi NN}}{m_\phi^2}n_S.
\end{aligned}
\end{equation}
We easily see that when hyperons are not entering, strangeness number density $n_S = 0$ which lead to mean field $\phi=0$. Also the gap equations for $\sigma$ at vacuum and finite density is 
\begin{equation}\label{eq_gap_sigma}
\begin{aligned}
\frac{\partial \Omega_H}{\partial \sigma}|_{{\rm vac}} &= -\bar{\mu}^2 f_\pi + \lambda_4 f_\pi^3 - \lambda_6 f_\pi^5 - m_\pi^2 f_\pi =0, \\
\frac{\partial \Omega_H}{\partial \sigma} &= - \bar{\mu}^2 \sigma + \lambda \sigma^3 - \lambda_6 \sigma^5 - m_\pi^2 f_\pi \\ 
& \quad + \frac{1}{2 \pi^2}\sum_{H} \sum_\alpha \sum_{i=\pm} m_{H, i} \frac{\partial m_{H, i} }{\partial \sigma} \left[ k_H^{i, \alpha} \hat{\mu}_H^{i, \alpha} \right. \\
&\quad \left.- m_{H, i}^2 \ln\left( \frac{k_H^{i, \alpha} + \hat{\mu}_H^{i, \alpha}}{m_{H, i}}\right) \right] =0
\end{aligned}
\end{equation}
with the mass derivative to $\sigma$ is
\begin{equation}
\begin{aligned}
\frac{\partial m_{N, \pm}}{ \partial \sigma} &= \frac{\partial m_{\Xi, \pm}}{ \partial \sigma} \\&=\frac{ (g_1 + g_2)^2 \sigma }{2\sqrt{(g_1 + g_2)^2 \sigma^2 + 4 m_0^2 }} \mp \frac{(g_1 - g_2)}{2}, \\
\frac{\partial m_{\Sigma, \pm}}{\partial \sigma} & = 0, \\
\frac{\partial m_{\Lambda, \pm}}{\partial \sigma} & = \frac{ (g_1 + g_2)^2  }{2\sqrt{(g_1 + g_2)^2 (\frac{4\sigma - \sigma_{s0}}{3})^2 + 4 m_0^2 }} \frac{4(4\sigma - \sigma_{s0})}{9} \\
&\quad \mp \frac{4(g_1 - g_2)}{6}.
\end{aligned}
\end{equation}

As discussed earlier, we treat $\sigma_s$ as a constant in the regime where hyperons are absent or have negligible density. Since the mass of $\Sigma$ baryons depends on $\sigma_s$ rather than $\sigma$ as in Eq.~(\ref{eq_mass_pm}), we have $\partial m_{\Sigma,\pm}/\partial \sigma = 0$. This means that $\Sigma$ baryons do not contribute to the $\sigma$-dependent terms in the gap equation~(\ref{eq_gap_sigma}), although they still contribute to the thermodynamic quantities through their mean-field contributions. With these considerations, the pressure of the system is obtained as

\begin{equation}
\begin{aligned}
P &= - \Omega_H  \\
&= - V(\sigma) + V(f_\pi) + \frac{1}{2}m_\omega^2 \omega^2 +  \frac{1}{2} m_\rho^2 \rho^2 \\
& \quad + \frac{1}{2}m_\phi^2 \phi^2 + \lambda_{\omega \rho} (g_{\omega NN} \omega)^2 (g_{\rho NN}\rho)^2 \\
& \quad + \frac{1}{\pi^2}\sum_H \sum_\alpha \sum_{i=\pm}\left[ \frac{1}{12}(k_H^{i, \alpha})^3 \hat{\mu}^{i, \alpha} - \frac{1}{8}m_{H, i}^2k_H^{i, \alpha} \hat{\mu}_H^{i, \alpha} \right. \\
& \quad \left.+ \frac{1}{8}m_{H, i}^4 \ln\left(\frac{k_H^{i, \alpha}  + \hat{\mu}_H^{i, \alpha}}{m_{H, i}}\right)   \right]
\end{aligned}
\end{equation}

For NS matter, we should include the leptonic contribution to obtain the total thermodynamic potential
\begin{align}
\Omega_{{\rm total}} = \Omega_H + \sum_{l = e, \mu} \Omega_l,
\end{align}
where $\Omega_l (l = e, \mu)$ are the thermodynamic potentials for leptons given as
\begin{align}
\Omega_l = -2 \int^{k_F} \frac{d^3 p}{(2 \pi)^3} (\mu_l - E_P^l).
\end{align}
We also need to consider the $\beta$ equilibrium and the charge neutrality conditions as
\begin{align}
\mu_e &= \mu_\mu = - \mu_Q, \\
\frac{\partial \Omega_{{\rm total}}}{\partial \mu_Q} &= \sum_{i=\pm} n^i_p + n^i_{\Sigma^+} - n^i_{\Sigma^-} - n^i_{\Xi^-} - n_l
\end{align}
where $\mu_Q$ is the charge chemical potential. The the total pressure in our hadronic model is 
\begin{align}
P_{{\rm total}} = - \Omega_{{\rm total}}.
\end{align}

\section{Numerical analysis}\label{sec_Numerical}

\subsection{Parameter determination}
In our model, for each choice of $m_0$, the four baryon sector parameters $g_1$, $g_2$, $m_1$, and $m_2$ are fully determined by fitting to the experimental baryon masses, as described in Section~\ref{sec_3_1}. This leaves 10 parameters  $\bar{\mu}^2$, $\lambda_4$, $\lambda_6$, $\lambda_8$, $g_{\omega NN}$, $g_{\rho NN}$, $g_{\phi NN}$, $\lambda_{\sigma \omega}$, $g_4$ and $\lambda_{\omega \rho}$ together with the meson masses. 

To fix these remaining parameters, we employ the following inputs and constraints: (1) the meson masses listed in Table~\ref{input: mass}, (2) the nuclear matter saturation properties given in Table~\ref{saturation}, (3) the gap equations for the chiral condensate $\sigma, \sigma_s$ at vacuum.  It is worth noting that in the density regime where hyperons have not yet appeared, the strangeness number density vanishes ($n_S = 0$), which implies that the mean field $\phi = 0$ according to Eq.~(\ref{eq_vector_gap}). Consequently, in this regime the coupling constant $g_{\phi NN}$ does not affect the system properties (e.g. the stiffness of EoS, onset density of hyperons) and can be determined independently after hyperons begin to appear at higher densities. However, since this study focuses primarily on the onset density of hyperons, we can simply set $g_{\phi NN}=0$. The determined parameters $\bar{\mu}^2$, $\lambda_4$, $\lambda_6$, $\lambda_8$, $g_{\omega NN}$, $g_{\rho NN}$ and $\lambda_{\omega \rho}$  for each choices of $m_0$ are listed in Table.~\ref{tab:parameters}. 
\begin{table}[htbp]
\centering
	\caption{  {\small Physical inputs in vacuum in unit of MeV.  }  }\label{input: mass}
	\begin{tabular}{cccccc}
		\hline\hline
		~$m_\pi$  ~&~ $m_{\phi}$ ~&~ $m_\omega$ ~&~ $m_\rho$ ~&~ $m_K$~\\
		\hline
		~140  ~&~ 1020 ~&~ 783 ~&~ 776~&~ 494 ~ \\
		\hline\hline
	\end{tabular}
\end{table}	
\begin{table}[htbp]
\centering
	\caption{  {\small Saturation properties used to determine the model parameters: the saturation density $n_0$, the binding energy $E_{\rm Bind}$, the incompressibility $K_0$, symmetry energy $S_0$, slope parameter $L_0$. 
 }  }
	\begin{tabular}{ccccc}\hline\hline
	~$n_0$ [fm$^{-3}$] ~& $E_{\rm Bind}$ [MeV] ~& $K_0$ [MeV] ~& $S_0$ [MeV] ~ & $L_0$ [MeV]\\
	\hline
	0.16 & 16 & 240 & 31 & 70  \\
	\hline\hline
	\end{tabular}
	\label{saturation}
\end{table}	
\begin{table}[h]
\caption{Values of parameters $\bar{\mu}^2, \lambda_4, \lambda_6, \lambda_8, g_{\omega NN}, g_{\rho NN}$ and $\lambda_{\omega \rho}$ for $m_0=500$--$900$ MeV.}
\label{tab:parameters}
\centering
\begin{tabular}{l|ccccc}
\hline\hline
~$m_0$ [MeV] ~~&~~ 500 ~~& ~~ 600 ~~&~~~ 700 ~~~& ~~~800 ~~~&~~~ 900 ~\\
\hline
$\bar{\mu}^2 / f^2_{\pi}$ & 75.17 & 65.83 & 51.28 & 33.17 & 12.23 \\

$\lambda_4$ & 186.21 & 167.85 & 137.37 & 97.92 & 49.19 \\

$\lambda_6 f^2_{\pi}$ & 157.36 & 146.16 & 125.89 & 98.36 & 61.79 \\

$\lambda_8 f^4_{\pi}$ & 3.03 & 2.89 & 2.63 & 2.24 & 1.69\\

$g_{\omega NN}$ & 7.74 & 6.93 & 6.05 & 4.96 & 3.23 \\

$g_{\rho NN}$ & 9.19 & 9.22 & 9.25 & 9.29 & 9.34 \\

$\lambda_{\omega \rho}$ & 0.07 & 0.10 & 0.16 & 0.34 & 1.74 \\
\hline\hline
\end{tabular}
\end{table}

\begin{figure}[htbp]
\centering
\includegraphics[width=0.5\hsize]{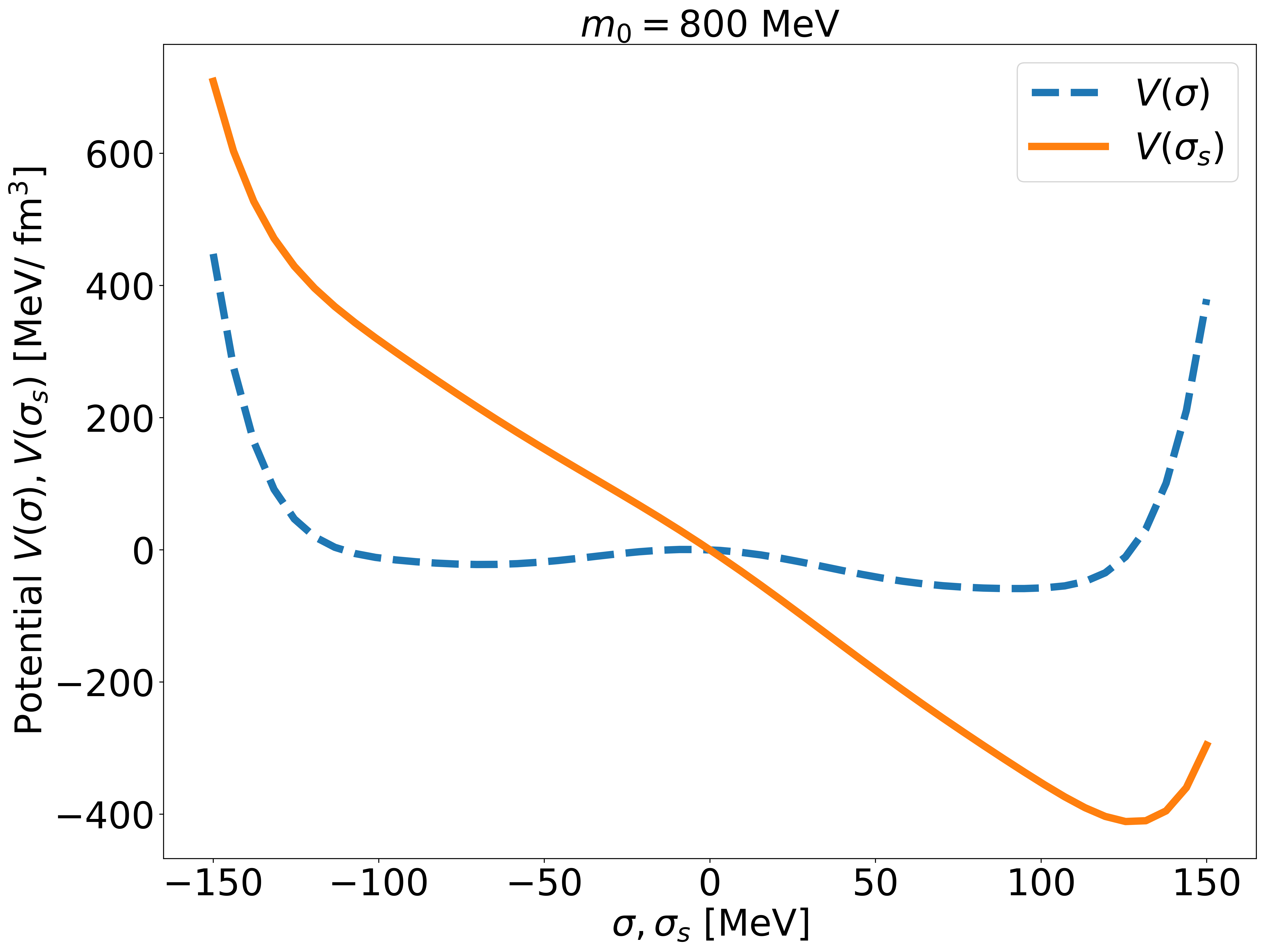}
\caption{Vacuum potential for $\sigma$ and $\sigma_s$ with $m_0$ fixed to be 800 MeV}
\label{fig_vac_potential}
\end{figure}

\begin{figure}[htbp]
\centering
\includegraphics[width=0.5\hsize]{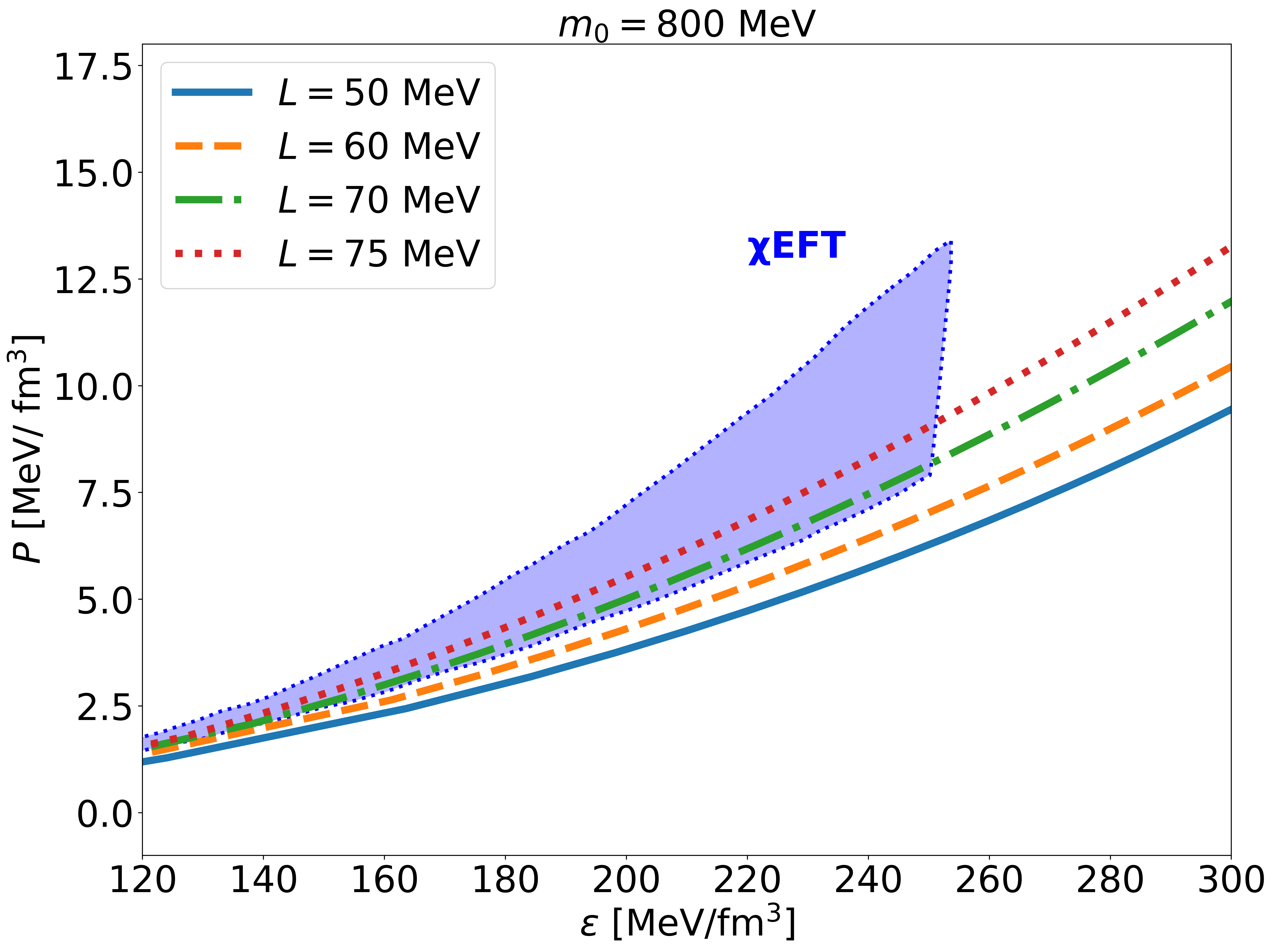}
\includegraphics[width=0.5\hsize]{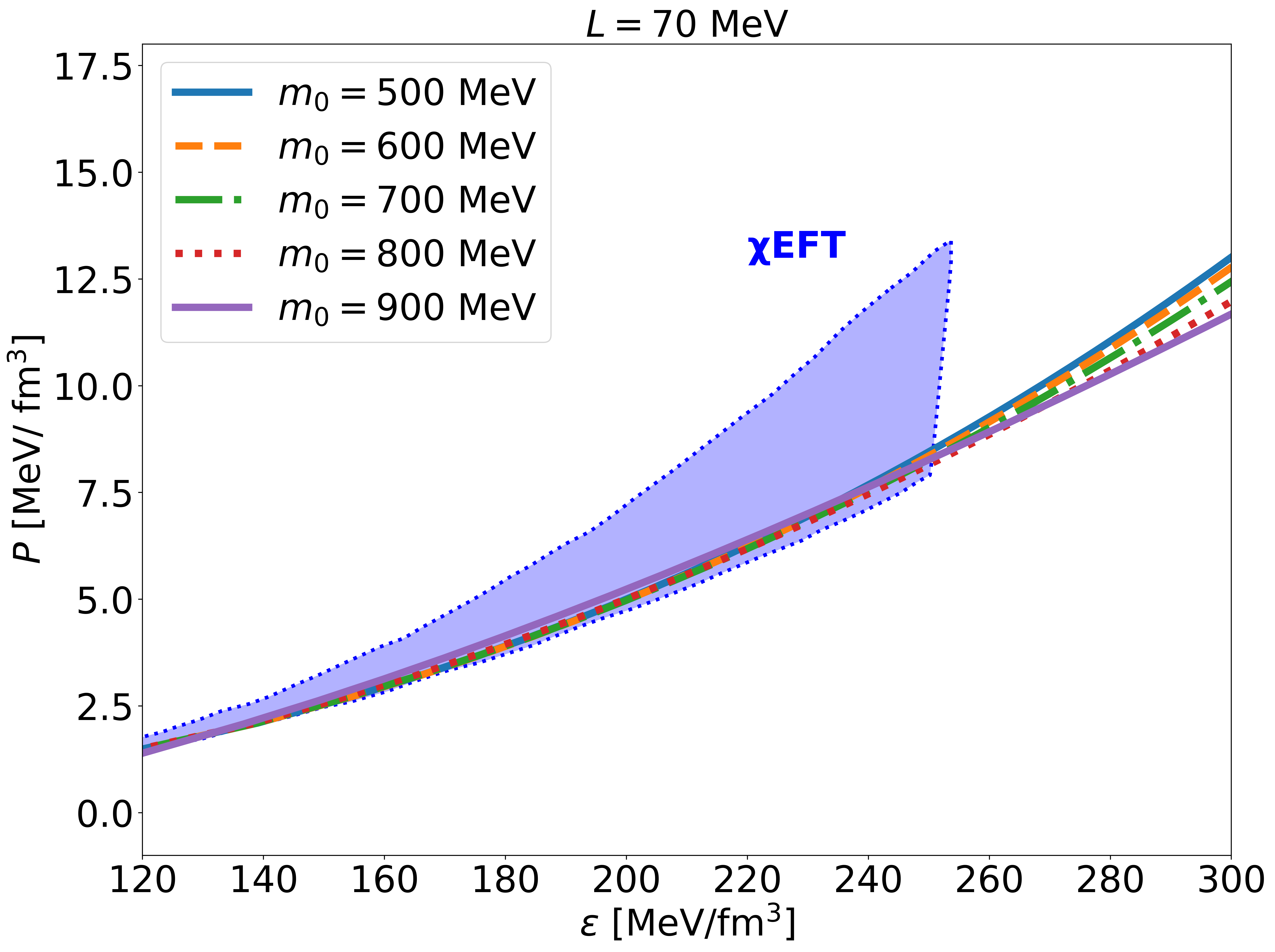}
\caption{EoS in the low-density regime. Upper panel:  $L = 50, 60, 70, 75$ MeV with fixed $m_0=800$ MeV. Lower panel: $m_0 = 500, 600, 700, 800, 900$ MeV with fixed $L=70$ MeV. The shaded blue band shows the  1$\sigma$ uncertainty range from $\chi$EFT~\cite{Drischler:2017wtt,Drischler:2020fvz,Keller:2022crb}. }
\label{fig_EOS}
\end{figure}

We then obtain the potential at zero density with the chiral invariant mass set to be $m_0 = 800~\text{MeV}$, as shown in Fig.~\ref{fig_vac_potential}. 
The potential develops a global minimum at $\sigma = f_\pi$ and $\sigma_s = 2f_K - f_\pi$, 
representing the physical vacuum where chiral symmetry is spontaneously broken and flavor SU(3) symmetry is explicitly broken. 
Figure~\ref{fig_EOS} presents the EoS in the low-density region before the onset of hyperons. 
In the upper panel of Fig.~\ref{fig_EOS}, we fix the chiral invariant mass at $m_0 = 800$ MeV and vary the value of slope parameter  $L_0$. For $L_0 = 50, 60$ MeV, the EoS is relatively soft---smaller pressure at a given energy density---and falls outside the calculations from chiral effective field theory ($\chi$EFT). As $L_0$ increases, the EoS becomes  stiffer and when $L_0 = 70$ MeV, the resulting EoS becomes consistent with $\chi$EFT calculations. In this study, we adopt $L_0 = 70$ MeV for our subsequent analysis, as listed in Table~\ref{saturation}, consistent with previous studies~\cite{universe7060182,Lattimer:2012xj,Li:2014oda}. In the lower panel of Fig.~\ref{fig_EOS}, we fix $L_0 = 70$ MeV and change the chiral invariant mass from $m_0 = 500$ to 900 MeV. All EoSs in this parameter range remain consistent with $\chi$EFT calculations~\cite{Drischler:2017wtt,Drischler:2020fvz,Keller:2022crb} in the low-density region. As the increasing of $m_0$ the EoS becomes softer as demonstrated in previous study of parity doublet model. 

\subsection{Hyperons in neutron star}

\begin{figure}[htbp]
\centering
\includegraphics[width=0.5\hsize]{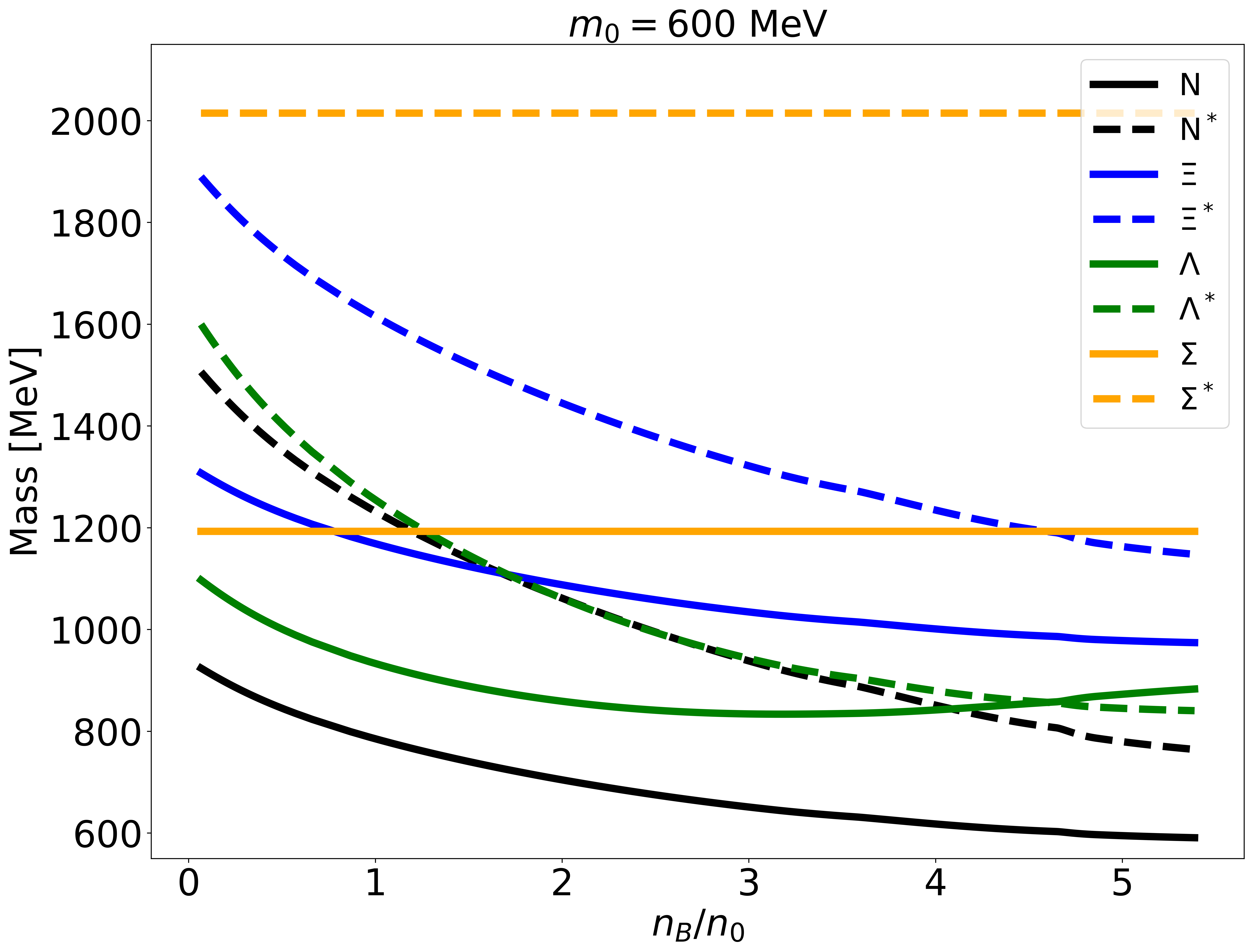}
\includegraphics[width=0.5\hsize]{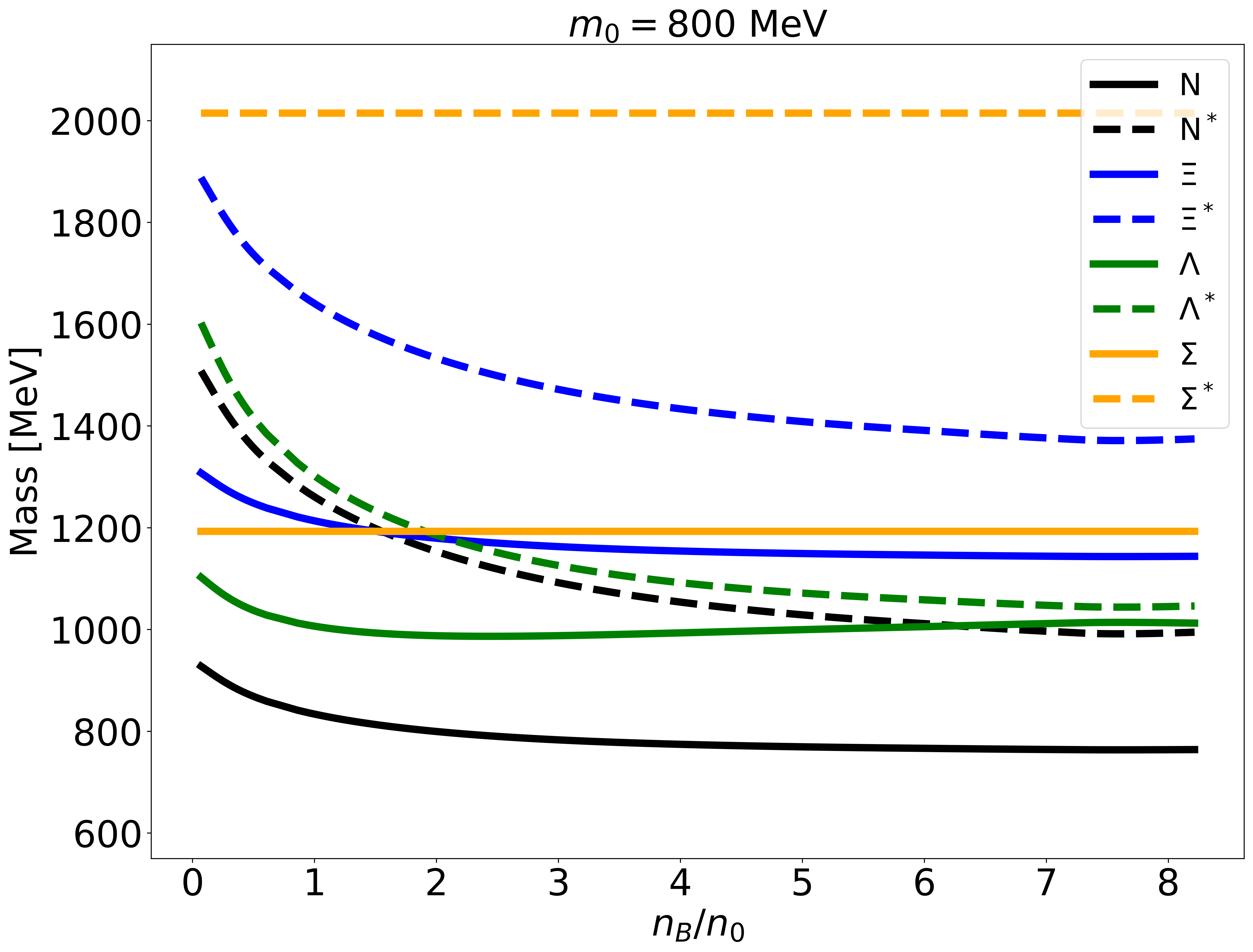}
\caption{Masses of $N, N^* \Xi, \Xi^*$ and $\Lambda, \Lambda^*$ as functions of the normalized baryon number density, with $m_0=600,  800$~MeV and $g_{\phi NN}=0$. The solid curves denote for the ground state mass while the dashed curves  denote for its excited state.  }
\label{fig_mass_nb}
\end{figure}

After determining all the parameters for each value of $m_0$, we analyze the density dependence of physical quantities. 
Figure~\ref{fig_mass_nb} presents the masses of $N$, $N^*$, $\Xi$, $\Xi^*$, $\Lambda$, and $\Lambda^*$ as functions of normalized baryon density $n_B/n_0$ for $m_0 = 600, 800$ MeV with $g_{\phi NN} = 0$. As discussed earlier, we treat $\sigma_s$ as constant and this assumption should be reliable in the density regime where hyperon densities are 0 or remain small values. In the case where we only consider  $(3, \bar{3}) + (\bar{3}, 3)$ representation, $\Sigma$ baryon masses remain constant as indicated in Eq.~(\ref{eq_mass_pm}).
%


In Fig.~\ref{fig_mass_nb}, as the increasing of density,  the mass splitting between ground states and their chiral partners for each baryon species become smaller, signaling partial restoration of the chiral symmetry. Also for small values of chiral invariant mass, the density dependence for $N, \Lambda, \Xi$ become stronger and have smaller masses at the density.

\begin{figure*}[htbp]
\centering
\includegraphics[width=1\hsize]{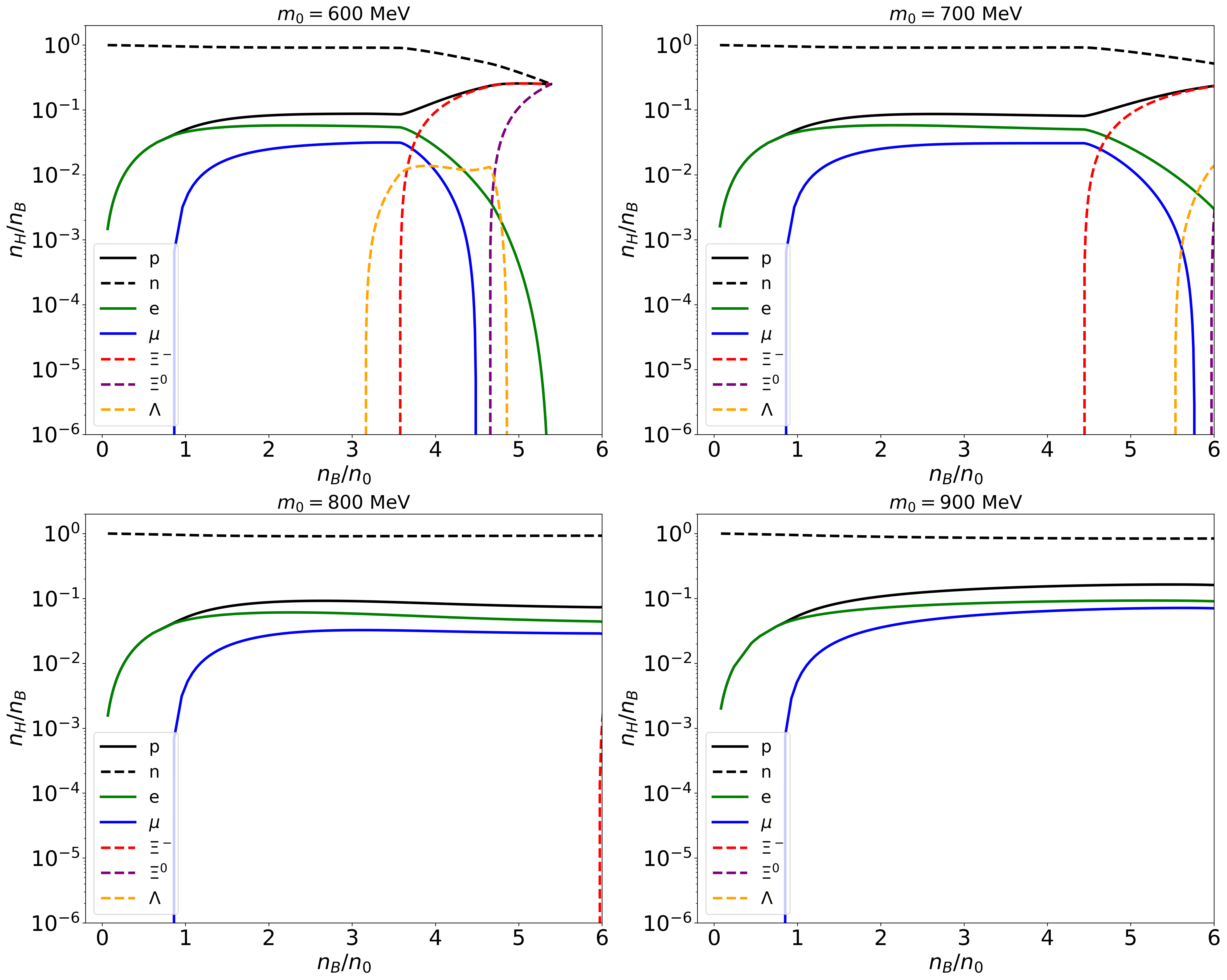}
\caption{Particle fractions $n_H/n_B$ as functions of normalized baryon density $n_B/n_0$ for different chiral invariant masses $m_0 = 600, 700, 800, 900$ MeV with $g_{\phi NN}=0$. The figure shows the evolution of neutrons (n), protons (p), hyperons ($\Xi^-, \Xi^0$), and leptons (e, $\mu$) in neutron star matter under $\beta$-equilibrium and charge neutrality conditions. The onset densities of hyperons increase significantly with larger $m_0$ values. }
\label{fig_ratio_nb}
\end{figure*}

When the effective chemical potential as defined in Eq~(\ref{eq_chemical_potential}) exceeds the corresponding baryon mass, that species begins populating the matter. 
Figure~\ref{fig_ratio_nb} displays the particle fraction $n_H/n_B$ for $m_0 = 600, 700, 800, 900$ MeV with $g_{\phi NN} = 0$. The muon first appears around $n_0$ for each choice of $m_0$, followed by hyperons ($\Xi^-$ and $\Xi^0$) at higher densities. The appearance of negatively charged $\Xi^-$ hyperons significantly impacts the leptonic composition. These hyperons contribute directly to charge neutrality, reducing the required lepton abundance. As $\Xi^-$ density increases, the electron and muon chemical potentials decrease to maintain charge balance. Since muons are much heavier than electrons ($m_\mu \approx 106$ MeV versus $m_e \approx 0.511$ MeV), they decouple first when their chemical potential drops below the muon mass threshold. This sequential lepton decoupling reorganizes the charged particle content, with important implications for neutrino opacity and NS cooling.

In Fig.~\ref{fig_ratio_nb}, we show the dependence of hyperon onset densities on the chiral invariant mass $m_0$. As $m_0$ increases from 600 to 900 MeV, the appearance of hyperons is  suppressed to higher densities. For $m_0 = 600$ MeV, the $\Lambda$ hyperon emerges at approximately $3.2n_0$ while $\Xi^-$ appears at $3.6n_0$. These onset densities shift dramatically to $5.5n_0$ and $4.5n_0$ respectively for $m_0 = 700$ MeV. At $m_0 = 800$ MeV, $\Xi^-$  enters at $6n_0$—approaching the densities correspond to those in NS cores—while $\Lambda$ hyperons are absent. For $m_0 = 900$ MeV, neither hyperon species appears within the density range considered.
This suppression offers an additional resolution to the hyperon puzzle. Since quark-hadron phase transition is expected to occur at densities between $2-5n_0$ and our results suggest that for sufficiently large $m_0$ values, matter undergoes a direct transition to quark matter before hyperons can populate. This scenario naturally prevents the EoS softening associated with hyperons.

We also note that the order in which hyperons appear is sensitive to the value of the chiral invariant mass $m_0$ for the $\Lambda$ and $\Xi$ hyperons. 
For small $m_0$, baryon masses originate predominantly from spontaneous chiral symmetry breaking, making them strongly dependent on the $\sigma$ field. 
As the baryon density increases and the non-strange condensate $\sigma$ decreases rapidly, baryons containing more non-strange quarks experience stronger mass reductions. 
Consequently, the $\Lambda$ hyperon, with its larger non-strange component, exhibits a faster mass reduction than the $\Xi$ hyperon, leading to its earlier onset in dense matter. 
In contrast, for large $m_0$, the weak $\sigma$-field dependence of baryon masses shifts the dominant mechanism from mass-driven to chemical-potential–driven hyperon emergence. 
As $m_0$ increases, the $\omega$-meson coupling becomes weaker while the $\rho$-meson coupling becomes relatively stronger. 
Then the contribution from $\hat{\mu}_S$ in Eq.~(\ref{eq_chemical_potential}) becomes less significant, while the negatively charged $\Xi^-$ hyperon gains an additional contribution from the isospin chemical potential. As a result, its effective chemical potential increases more rapidly with density than that of the neutral $\Lambda$, leading to an earlier onset of the $\Xi^-$ hyperon.
The chiral invariant mass $m_0$ serves as a key parameter governing not only whether hyperons appear in neutron-star matter but also the underlying mechanism of their emergence.

\begin{figure}[htbp]
\centering
\includegraphics[width=0.5\hsize]{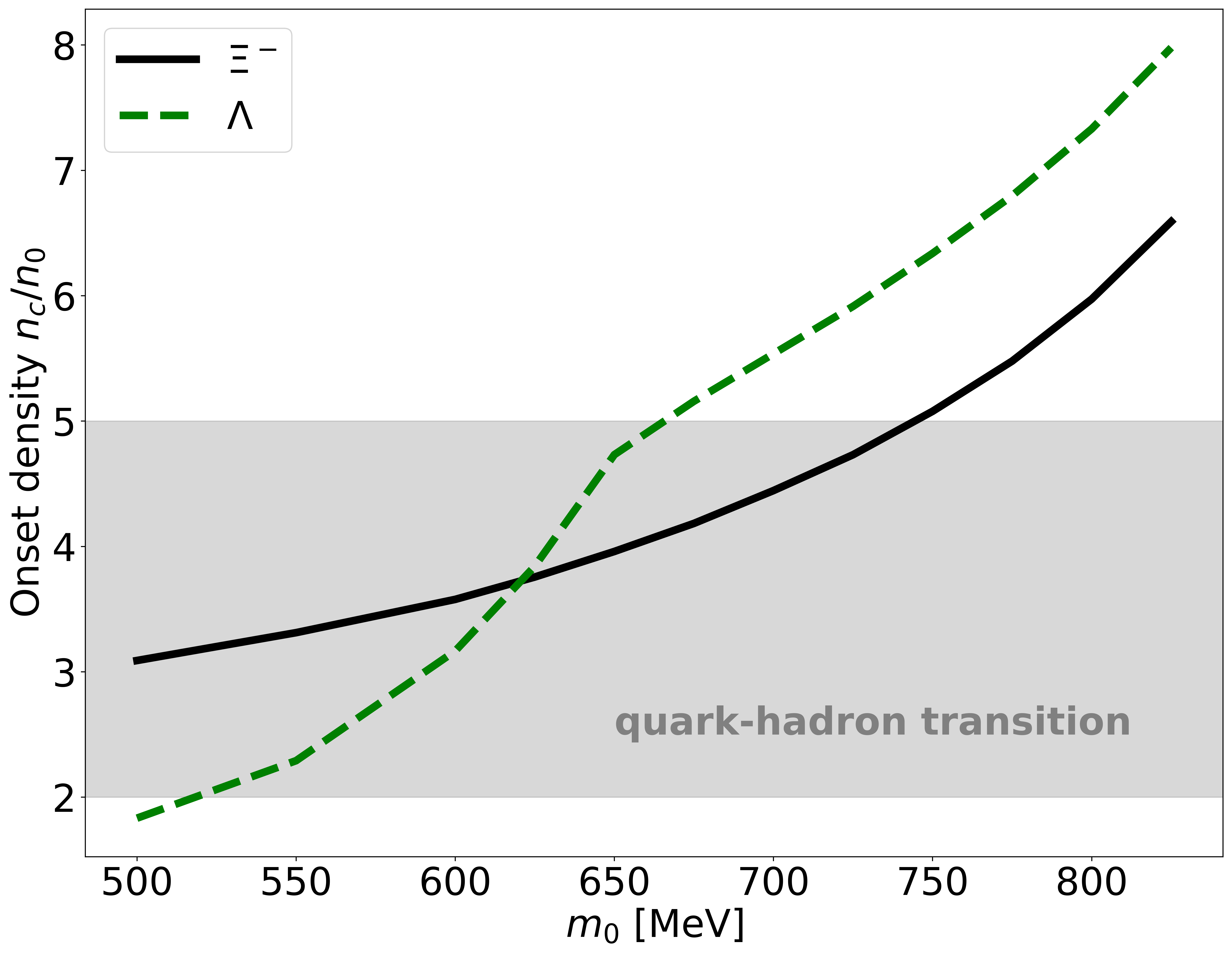}
\caption{Onset density of hyperon $n_c / n_0$ as the function of chiral invariant mass $m_0$ with $g_{\phi NN}=0$. }
\label{fig_onset_density}
\end{figure}

In Figure~\ref{fig_onset_density}, we show the onset densities of the $\Lambda$ and $\Xi^-$ hyperons as functions of the chiral invariant mass $m_0$. 
For $m_0 \lesssim 625~\mathrm{MeV}$, the $\Xi^-$ hyperon appears at a higher density than the $\Lambda$.
As $m_0$ increases beyond approximately $625~\mathrm{MeV}$, this trend reverses, and the $\Xi^-$ begins to emerge earlier than the $\Lambda$. 
In addition, the overall onset densities of both hyperons increase with $m_0$. 
When $m_0$ exceeds about $675~\mathrm{MeV}$, the onset of the $\Lambda$ hyperon lies above the density range associated with the quark--hadron transition, and for the $\Xi^-$ this occurs at around $750~\mathrm{MeV}$. 
Consequently, hyperons effectively decouple from neutron star matter at large $m_0$, thereby preventing the softening of the EoS due to hyperon formation.


Finally, we note that the hyperon potentials at saturation density, defined as
\begin{align}\label{eq_Ypotential}
U_H = m_H(\sigma)|_{\rm sat.} - m_H(f_\pi) + g_{\omega H}\omega, \quad H = \Lambda, \Sigma, \Xi,
\end{align}
are not reproduced in our construction. In the present framework, the nucleon-sector parameters are entirely determined by the nuclear saturation properties listed in Table~\ref{saturation}, while the hyperon-vector meson couplings follow from the hidden local symmetry, consistent with the SU(6) quark model~\cite{Schaffner:1995th} relations:
\begin{align}
\frac{1}{2}g_{\omega \Sigma\Sigma} = \frac{1}{2}g_{\omega \Lambda\Lambda} = g_{\omega \Xi\Xi} = \frac{1}{3}g_{\omega NN}.
\end{align}
Consequently, there are no additional free parameters available to adjust the hyperon potential depths and our model predictions is presented in Table~\ref{tab:hyperon_potential}. Our predicted hyperon potentials for $\Lambda, \Xi$ are deeper (more attractive) than the empirical values ( $U_\Lambda = -30$ MeV, $U_{\Xi}$=-20 MeV ). Since deeper potentials lead to earlier hyperon onset, correctly reproducing the experimental potentials would delay the onset densities even further. Even when the hyperon potentials are correctly reproduced, a sufficiently large chiral invariant mass still naturally suppresses hyperon appearance.

\begin{table}[h]
\caption{Predicted values of hyperon potential for $m_0=500$--$900$ MeV.}
\label{tab:hyperon_potential}
\centering
\begin{tabular}{lccccc}
\hline\hline
~$m_0$ [MeV] ~&~~ 500 ~~&~~ 600 ~~&~~~ 700 ~~~& ~~~800 ~~~&~~~ 900 ~~~ \\
\hline
$U_\Lambda$ [MeV] & -141.87 & -120.19 & -99.31 & -77.03 & -48.94 \\ 
$U_\Sigma$  [MeV]& 80 & 64.15& 48.98 & 32.88 & 13.98 \\ 
$U_\Xi$ [MeV] & -133.81 & -118.73 & -104.41 & -88.61 & -68.93  \\ 
\hline\hline
\end{tabular}
\end{table}

To address the hyperon potential discrepancy within the $(3, \bar{3}) + (\bar{3}, 3)$ representation, , we clarify the fundamental limitations inherent to this chiral representation. As demonstrated in Ref.~\cite{Gao:2025eax}, the $(3, \bar{3}) + (\bar{3}, 3)$ representation alone produces an unphysical degeneracy $m_N = m_\Xi$ due to the antisymmetric tensor structure in the Yukawa interactions, which prevents the strange quark condensate $\sigma_s$ from contributing to the nucleon and $\Xi$ baryon masses. Furthermore, this representation generates only D-type interactions, while the F-type interactions essential for properly describing the baryon mass spectrum and hyperon-meson couplings are absent. Consequently, even with explicit breaking terms $m_1, m_2$, the model cannot reproduce the hyperon potentials and the mass spectrum simultaneously. To achieve a quantitative reproduction of both the hyperon mass spectrum and the potential depths, future work should incorporate higher chiral representations, such as the $(3, 6) + (6, 3)$ channels containing ``bad'' diquark configurations~\cite{Gao:2025eax}. Including these representations would introduce additional Yukawa couplings and mixing effects that modify the density-dependent mass behavior of $\Sigma$, $\Xi$, and $\Lambda$ hyperons. Alternatively, one may relax the constraint $a_V = a_V^\prime$ for the vector meson couplings, which would allow mixing between the singlet $\Lambda^0$ and octet $\Lambda^8$ states, as demonstrated in Appendix~\ref{sec_ap1}. This modification introduces additional freedom to adjust the  hyperon potential independently while preserving the overall chiral structure of the model.
While the specific onset densities would shift depending on the specific modification, the fundamental mechanism we have identified---the role of the chiral invariant mass $m_0$ in controlling hyperon appearance---remains robust, since the underlying mechanism is governed by the weakened density dependence of baryon masses at large $m_0$, which is a generic feature of the parity doublet structure independent of the specific hyperon potential values. If $m_0$ is sufficiently large, hyperons appear only at densities where quark degrees of freedom are expected to dominate, or perhaps not at all within realistic neutron star cores. The transition from hadronic to quark matter at $2$--$5n_0$ then prevents the problematic EoS softening traditionally associated with hyperons.




\section{SUMMARY AND DISCUSSIONS}
\label{summary}

In this work, we have investigated the appearance of hyperons in NS matter within the framework of the linear SU(3) parity doublet model. By extending the conventional SU(2) parity doublet approach to include the complete baryon octet, we have developed a description of dense matter that naturally incorporates chiral symmetry restoration and its effects on hyperon dynamics.

Our approach is based on the chiral representation $(3,\bar{3}) + (\bar{3},3)$, which emerges naturally from the quark-diquark picture where ground state baryons are dominated by ``good'' diquark correlations in the flavor-antisymmetric $\bar{3}$ representation. This choice of representation ensures that our effective theory captures the essential physics of baryon structure while maintaining consistency with QCD symmetries. The model parameters were determined by fitting to vacuum baryon masses and nuclear saturation properties, with the chiral invariant mass $m_0$ emerging as the key parameter controlling hyperon behavior in dense matter.

The principal finding of this study is that the onset density of hyperons have strong dependence on the chiral invariant mass $m_0$. For $m_0 = 500~\mathrm{MeV}$, the $\Lambda$ and $\Xi^-$ hyperons appear at around $1.9n_0$ and $3n_0$, respectively. As $m_0$ increases, the onset densities shift significantly, and for $m_0 \gtrsim 750~\mathrm{MeV}$, both hyperons emerge only above $5n_0$. This tendancy originates from the interplay between chiral symmetry restoration and the mass generation mechanism inherent in the parity doublet model. Importantly, this qualitative behavior persists even when additional chiral representations such as $(3,6) + (6,3)$ or $(8,1) + (1,8)$ are incorporated, since the presence of the chiral invariant mass $m_0$ consistently weakens the density dependence of baryon masses and delays hyperon onset.
This strong $m_0$ dependence offers an additional solution for the hyperon puzzle. If the hyperon onset density exceeds the typical range of the quark–hadron phase transition ($2\text{--}5n_0$), the system undergoes deconfinement before hyperons can appear. In this case, the EoS avoids the excessive softening caused by hyperon formation while remaining compatible with the existence of massive neutron stars.

\begin{figure}[htbp]
\centering
\includegraphics[width=0.5\hsize]{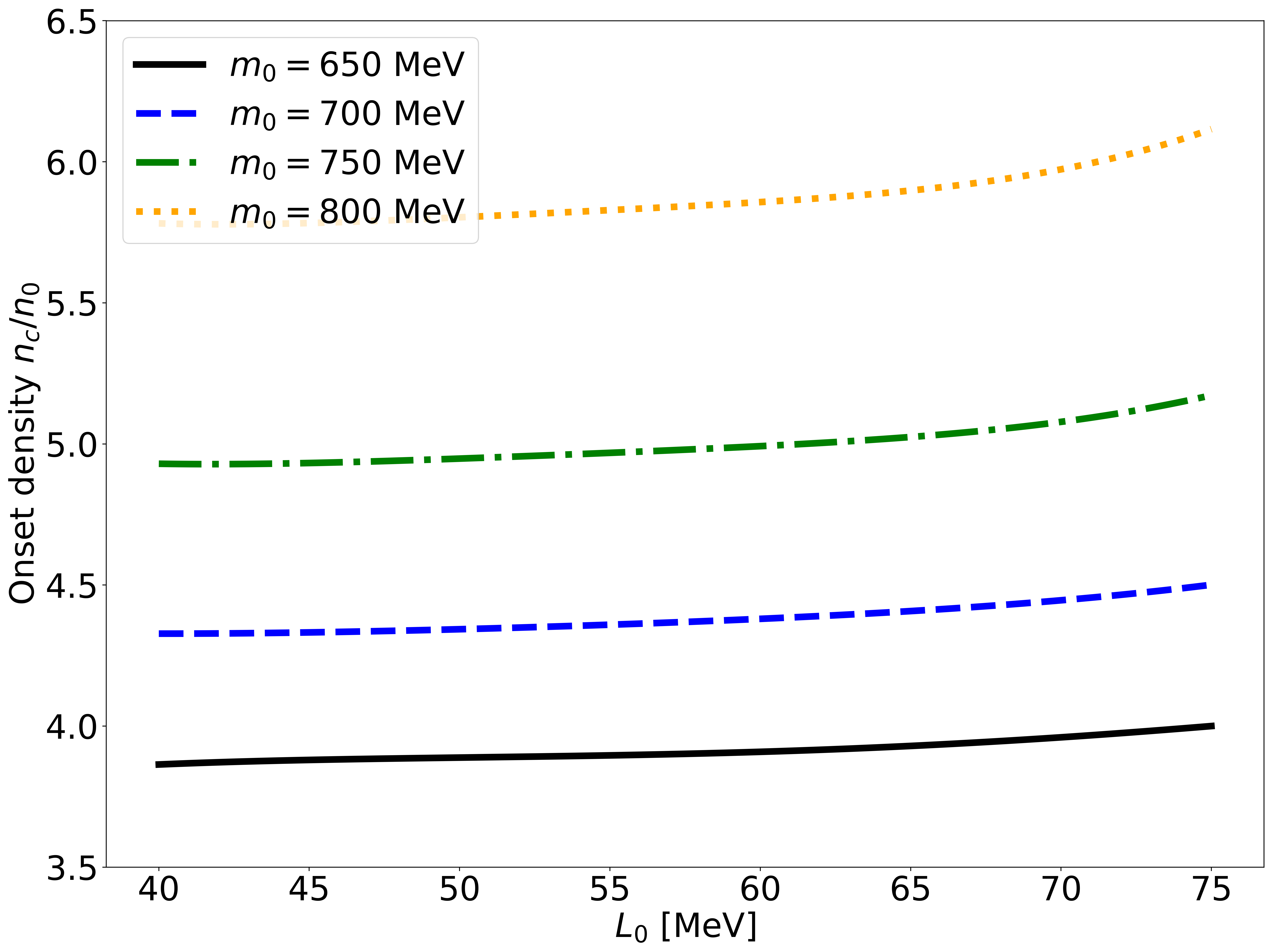}
\caption{Onset density of  $\Xi^-$ hyperon  as the function of slope parameter $L_0$ for several choices of chiral invariant mass $m_0$ with $g_{\phi NN}=0$. }
\label{fig_onset_L}
\end{figure}

\begin{figure}[htbp]
\centering
\includegraphics[width=0.5\hsize]{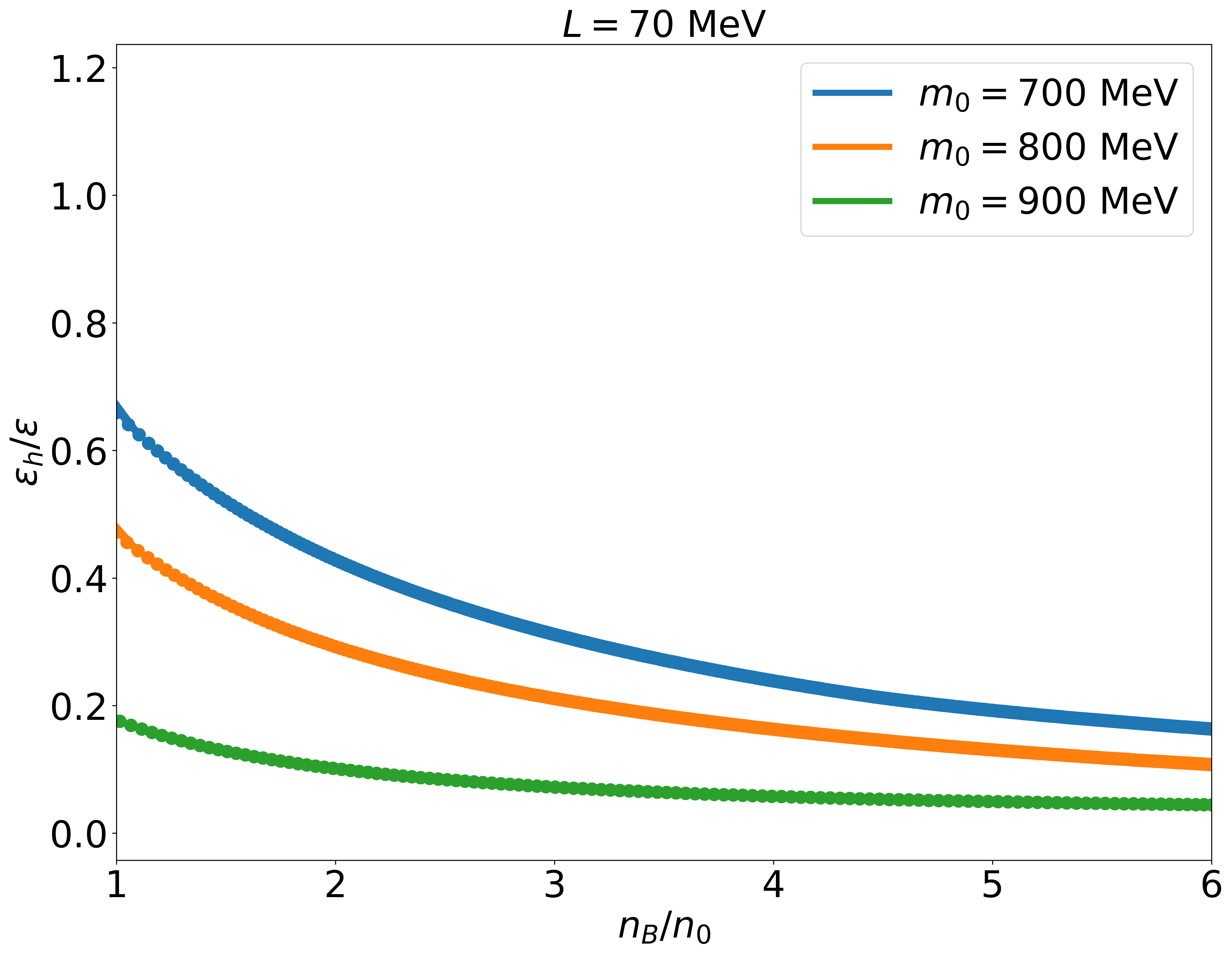}
\caption{Ratio of the energy density from higher-order terms $\varepsilon_h$ to the total energy density $\varepsilon$ as a function of normalized baryon density $n_B/n_0$ for $m_0 = 700$, $800$, and $900$ MeV with $L_0 = 70$ MeV. The higher-order contributions include terms from the scalar potential with coefficients $\lambda_{4\sim 8}$ and the $\omega^2\rho^2$ coupling term proportional to $\lambda_{\omega\rho}$. }
\label{fig_higher_order}
\end{figure}

We also note that the choice of the slope parameter $L_0$ influences the onset density of hyperons. Variations in $L_0$ modify the $\rho$-meson coupling, thereby affecting the effective baryon chemical potentials. Throughout this study, we adopt $L_0 = 70$ MeV to remain consistent with $\chi$EFT calculations and other theoretical studies. Nonetheless, a phenomenological analysis can still be performed for different values of $L_0$. In Fig.~\ref{fig_onset_L}, we present the onset density of the $\Xi^-$ hyperon as a function of $L_0$ for several choices of the chiral invariant mass $m_0$. As $L_0$ increases, the onset density of $\Xi^-$ exhibits only a slight increase, indicating that while the slope parameter has a minor effect, the dominant factor governing the hyperon onset remains the chiral invariant mass $m_0$.
It is also important to examine the role of higher-order terms in our model. The scalar potential $V_M$ contains terms with coefficients $\lambda_{4\sim 8}$ , and the $\omega$--$\rho$ coupling proportional to $\lambda_{\omega\rho}$  effectively generates many-body baryon interactions beyond the two-body level. To quantify their contributions, we calculate the ratio $\varepsilon_h/\varepsilon$, where $\varepsilon_h$ denotes the energy density from all higher-order terms and $\varepsilon$ is the total energy density. Figure~\ref{fig_higher_order} shows this ratio as a function of the normalized baryon density $n_B/n_0$ for several values of the chiral invariant mass $m_0$. As  density increases toward the regime relevant for neutron star cores ($n_B/n_0 \gtrsim 3$), the ratio $\varepsilon_h/\varepsilon$ decreases substantially for all $m_0$ values, falling below $20\%$ at $n_B/n_0 \sim 5$--$6$. This behavior indicates that while the higher-order terms provide non-negligible contributions near saturation density, their relative importance diminishes at high densities where hyperon onset and the stiffening of the equation of state become relevant. The dominant physics governing the EoS at high densities is therefore controlled by the two-body interactions and the chiral invariant mass $m_0$, supporting our conclusion that $m_0$ serves as the key mechanism for resolving the hyperon puzzle without requiring additional \textit{ad hoc} repulsive hyperon interactions.

Several important directions for future work should be noted. First, our analysis has focused on the $(3,\bar{3}) + (\bar{3},3)$ representation, which captures the dominant ``good'' diquark contributions. A more complete treatment would include the $(8, 1) + (1, 8)$ and $(3,6) + (6,3)$ chiral representations, which could modify the detailed density dependence of baryon masses, particularly for $\Sigma$ and $\Lambda$ hyperons. Second, we have treated the strange condensate $\sigma_s$ as constant, justified in the low hyperon density regime but requiring revision for a full description of high-density matter. Third, the coupling constant $g_{\phi NN}$ remains undetermined in our current framework and requires additional constraints from hypernuclear physics or lattice QCD.

Future investigations should also explore the implications of our findings for NS observables beyond the mass-radius relationship. The delayed or absent hyperon onset would significantly impact neutrino opacity and cooling rates, potentially resolving tensions between theoretical cooling curves and observational data. Additionally, the modified composition would affect transport properties relevant to NS mergers and gravitational wave signatures.

\section*{Acknowledgement}
The author gratefully acknowledge helpful conversations with Atsushi Hosaka, Masayasu Harada and Toru Kojo.

\appendix
\section{Detailed calculations}\label{sec_ap1}
In this section, we show the detailed calculation of Lagrangian Eq.~(\ref{eq_vector_B_L}). The coefficients of the coupling for $\psi_1$ and $\psi_2$ should be same and we obtain the couplings for each components of $V^\mu$ field. For the (1, 1) component, we set $V^{\mu}=V_L^{\mu}=V_R^\mu=\delta_\mu^0 {\rm diag}(1, 0, 0)$
\begin{equation}
\begin{aligned}
\mathcal{L}_B^{\mathrm{Vec}}= & -g a_{V} \operatorname{tr} \psi_1^{\dagger} \operatorname{diag}(1,0,0) \psi_1 \\
= & -g a_{V}\left[p^{\dagger} p+\left(\Sigma^{+}\right)^{\dagger} \Sigma^{+}+\frac{1}{2}\left(\Sigma^0\right)^{\dagger} \Sigma^0\right.\\
&\left.+\frac{1}{3}\left(\Lambda^0\right)^{\dagger} \Lambda^0+\frac{1}{6}\left(\Lambda^8\right)^{\dagger} \Lambda^8\right] \\
& -g a_{V}^{\prime}\left[p^{\dagger} p+\left(\Sigma^{+}\right)^{\dagger} \Sigma^{+}+\frac{1}{2}\left(\Sigma^0\right)^{\dagger} \Sigma^0+\frac{2}{3}\left(\Lambda^0\right)^{\dagger} \Lambda^0\right.\\
&\left.+\frac{5}{6}\left(\Lambda^8\right)^{\dagger} \Lambda^8+n^{\dagger} n+\left(\Xi^0\right)^{\dagger} \Xi^0\right] \\
& -g\left(a_{V}-a_{V}^{\prime}\right)\left[\frac{1}{3 \sqrt{2}}\left(\Lambda^0\right)^{\dagger} \Lambda^8+\frac{1}{\sqrt{6}}\left(\Lambda^0\right)^{\dagger} \Sigma^0\right.\\
&\left.+\frac{1}{2 \sqrt{3}}\left(\Lambda^8\right)^{\dagger} \Sigma^0+\text { h.c. }\right]
\end{aligned}
\end{equation}
For the (2, 2) component, $V_L^{\mu}=V_R^\mu=\delta_\mu^0 {\rm diag}(0, 1, 0)$
\begin{equation}
\begin{aligned}
\mathcal{L}_B^{\mathrm{Vec}}= & -g a_{V} \operatorname{tr} \psi_1^{\dagger} \operatorname{diag}(0,1,0) \psi_1\\
&-g a_{V}^{\prime} \operatorname{tr} \psi_1^{\dagger} \psi_1(1-\operatorname{diag}(0,1,0)) \\
= & -g a_{V}\left[n^{\dagger} n+\left(\Sigma^{-}\right)^{\dagger} \Sigma^{-}+\frac{1}{2}\left(\Sigma^0\right)^{\dagger} \Sigma^0 \right.\\
&\left.+\frac{1}{3}\left(\Lambda^0\right)^{\dagger} \Lambda^0+\frac{1}{6}\left(\Lambda^8\right)^{\dagger} \Lambda^8\right] \\
& -g a_{V}^{\prime}\left[n^{\dagger} n+\left(\Sigma^{-}\right)^{\dagger} \Sigma^{-}+\frac{1}{2}\left(\Sigma^0\right)^{\dagger} \Sigma^0+\frac{2}{3}\left(\Lambda^0\right)^{\dagger} \Lambda^0\right.\\
&\left.+\frac{5}{6}\left(\Lambda^8\right)^{\dagger} \Lambda^8+p^{\dagger} p+\left(\Xi^{-}\right)^{\dagger} \Xi^{-}\right] \\
& -g\left(a_{V}-a_{V}^{\prime}\right)\left[\frac{1}{3 \sqrt{2}}\left(\Lambda^0\right)^{\dagger} \Lambda^8-\frac{1}{\sqrt{6}}\left(\Lambda^0\right)^{\dagger} \Sigma^0\right.\\
&\left.-\frac{1}{2 \sqrt{3}}\left(\Lambda^8\right)^{\dagger} \Sigma^0+\text { h.c. }\right]+ {\rm Mirror} .
\end{aligned}
\end{equation}
For the (3, 3) component, $V_L^{\mu}=V_R^\mu=\delta_\mu^0 {\rm diag}(0, 0, 1)$
\begin{equation}
\begin{aligned}
\mathcal{L}_B^{\mathrm{Vec}}=&-g a_{V} \operatorname{tr} \psi_1^{\dagger} \operatorname{diag}(0,0,1) \psi_1\\
&-g a_{V}^{\prime} \operatorname{tr} \psi_1^{\dagger} \psi_1(1-\operatorname{diag}(0,0,1))\\
= & -g a_{V}\left[\left(\Xi^0\right)^{\dagger} \Xi^0+\left(\Xi^{-}\right)^{\dagger} \Xi^{-}\right.\\
&\left.+\frac{1}{3}\left(\Lambda^0\right)^{\dagger} \Lambda^0+\frac{2}{3}\left(\Lambda^8\right)^{\dagger} \Lambda^8\right] \\
& -g a_{V}^{\prime}\left[\left(\Xi^0\right)^{\dagger} \Xi^0+\left(\Xi^{-}\right)^{\dagger} \Xi^{-}+\frac{2}{3}\left(\Lambda^0\right)^{\dagger} \Lambda^0+\frac{1}{3}\left(\Lambda^8\right)^{\dagger} \Lambda^8\right.\\
&\left.+\left(\Sigma^{+}\right)^{\dagger} \Sigma^{+}+\left(\Sigma^{-}\right)^{\dagger} \Sigma^{-}+\left(\Sigma^0\right)^{\dagger} \Sigma^0\right] \\
& -g\left(a_{V}-a_{V}^{\prime}\right)\left[-\frac{2}{3 \sqrt{2}}\left(\Lambda^0\right)^{\dagger} \Lambda^8+\text { h.c. }\right]+ {\rm Mirror}.
\end{aligned}
\end{equation}
If $a_V = a_V^{\prime}$ and we take the whole expression as $V_L^{\mu}=V_R^\mu=\delta^\mu_0 {\rm diag}((\omega + \rho)/2, (\omega - \rho)/2, \phi / \sqrt{2})$, then 
\begin{equation}
\begin{aligned}
\mathcal{L}_B^{{\rm Vec}} =& -g a_{V} \operatorname{tr}\left[\bar{\psi}_{1 L} \gamma_\mu V_L^\mu \psi_{1 L}\right. \\
&\left.+\bar{\psi}_{1 L} \gamma_\mu \psi_{1 L}\left(\operatorname{tr} V_R^\mu-V_R^\mu\right)+(L \leftrightarrow R)\right]\\
= & -g a_{V}\{ \\
& +\frac{\omega}{2} \operatorname{tr}\left[3 p^{\dagger} p+3 n^{\dagger} n+2\left(\Sigma^{+}\right)^{\dagger} \Sigma^{+}+2\left(\Sigma^{-}\right)^{\dagger} \Sigma^{-}\right.\\
&\left.+2\left(\Sigma^0\right)^{\dagger} \Sigma^0+2\left(\Lambda^0\right)^{\dagger} \Lambda^0+2\left(\Lambda^8\right)^{\dagger} \Lambda^8 \right.\\
&\left.+\left(\Xi^0\right)^{\dagger} \Xi^0+\left(\Xi^{-}\right)^{\dagger} \Xi^{-}\right] \\
& +\frac{\rho}{2} \operatorname{tr}\left[p^{\dagger} p-n^{\dagger} n+2\left(\Sigma^{+}\right)^{\dagger} \Sigma^{+}-2\left(\Sigma^{-}\right)^{\dagger} \Sigma^{-} \right.\\
&\left.+\left(\Xi^0\right)^{\dagger} \Xi^0-\left(\Xi^{-}\right)^{\dagger} \Xi^{-}\right] \\
& \left. +\frac{\phi}{\sqrt{2}} \operatorname{tr}\left[\left(\Sigma^{+}\right)^{\dagger} \Sigma^{+}+\left(\Sigma^{-}\right)^{\dagger} \Sigma^{-}+\left(\Sigma^0\right)^{\dagger} \Sigma^0+\left(\Lambda^0\right)^{\dagger} \Lambda^0 \right. \right.\\
&\left. \left.+\left(\Lambda^8\right)^{\dagger} \Lambda^8+2\left(\Xi^0\right)^{\dagger} \Xi^0+2\left(\Xi^{-}\right)^{\dagger} \Xi^{-}\right]\right\}
\end{aligned}
\end{equation}
We then finally obtain the magnitudes of each interaction between the vector mesons and the baryons correspond to the definition of the effective baryon chemical potential in Eq.~(\ref{eq_chemical_potential_0}).

\bibliography{ref_3fPDM_2022.bib}

\end{document}